\documentclass{sig-alternate-2013}

\newfont{\mycrnotice}{ptmr8t at 7pt}
\newfont{\myconfname}{ptmri8t at 7pt}
\let\crnotice\mycrnotice%
\let\confname\myconfname%

\usepackage{times}
\usepackage{verbatim}
\usepackage{bm}
\usepackage{color}
\usepackage{balance}
\makeatletter
\newif\if@restonecol
\makeatother

\usepackage[lined,algonl,boxed]{algorithm2e}
\usepackage{epsfig}
\usepackage{epstopdf}
\usepackage{graphicx}
\usepackage{subfigure}
\usepackage{multirow}
\usepackage{amsmath}
\usepackage{amssymb}
\usepackage[bookmarks=true]{hyperref}

\newcommand{\hide}[1]{} 
\newcommand{\vpara}[1]{\vspace{0.03in}\noindent\textbf{#1 }}

\newdef{definition}{Definition}
\newdef{problem}{Problem}
\newdef{theorem}{Theorem}
\newdef{lemma}{Lemma}

\def\hindex{\textit{h}-index}
\def\hindices{\textit{h}-indices}
\def\fonescore{$F_1$ score}

\permission{Permission to make digital or hard copies of all or part of this work for personal or classroom use is granted without fee provided that copies are not made or distributed for profit or commercial advantage and that copies bear this notice and the full citation on the first page. Copyrights for components of this work owned by others than ACM must be honored. Abstracting with credit is permitted. To copy otherwise, or republish, to post on servers or to redistribute to lists, requires prior specific permission and/or a fee. Request permissions from Permissions@acm.org.}
\conferenceinfo{WSDM 2015,}{February 02 - 06 2015, Shanghai, China}
\copyrightetc{Copyright 2015 ACM \the\acmcopyr}
\crdata{978-1-4503-3317-7/15/02\ ...\$15.00\\
http://dx.doi.org/10.1145/2684822.2685314}
\begin{document}

\title{Will This Paper Increase Your {\huge\textit{h}}-index? \\ Scientific Impact Prediction}

\numberofauthors{3}
\author{
\alignauthor Yuxiao Dong \\
\affaddr{Dept. of Computer Science and Engineering, and iCeNSA} \\
\affaddr{University of Notre Dame} \\
\affaddr{Notre Dame, IN 46556}
\email{ydong1@nd.edu}
\alignauthor Reid A. Johnson \\
\affaddr{Dept. of Computer Science and Engineering, and iCeNSA} \\
\affaddr{University of Notre Dame} \\
\affaddr{Notre Dame, IN 46556}
\email{rjohns15@nd.edu}
\alignauthor Nitesh V. Chawla \\
\affaddr{Dept. of Computer Science and Engineering, and iCeNSA} \\
\affaddr{University of Notre Dame} \\
\affaddr{Notre Dame, IN 46556}
\email{nchawla@nd.edu}
}
\maketitle
 \sloppy
\begin{abstract}

Scientific impact plays a central role in the evaluation of the output of scholars, departments, and institutions.
A widely used measure of scientific impact is citations, with a growing body of literature focused on predicting the number of citations obtained by any given publication.
The effectiveness of such predictions, however, is fundamentally limited by the power-law distribution of citations, whereby publications with few citations are extremely common and publications with many citations are relatively rare.
Given this limitation, in this work we instead address a related question asked by many academic researchers in the course of writing a paper, namely: ``Will this paper increase my \hindex?''
Using a real academic dataset with over 1.7 million authors, 2 million papers, and 8 million citation relationships from the premier online academic service ArnetMiner, we formalize a novel scientific impact prediction problem to examine several factors that can drive a paper to increase the primary author's \hindex.
We find that the researcher's authority on the publication topic and the venue in which the paper is published are crucial factors to the increase of the primary author's \hindex, while the topic popularity and the co-authors' \hindices\ are of surprisingly little relevance.
By leveraging relevant factors, we find a greater than 87.5\% potential predictability for whether a paper will contribute to an author's \hindex\ within five years.
As a further experiment, we generate a self-prediction for \textit{this} paper, estimating that there is a 76\% probability that it will contribute to the \hindex\ of the co-author with the highest current \hindex\ in five years.
We conclude that our findings on the quantification of scientific impact can help researchers to expand their influence and more effectively leverage their position of ``standing on the shoulders of giants.''

\end{abstract}

{\small
\category{H.2.8}{Database Management}{Database Applications---\textit{Data Mining}} 
\category{H.3.7}{Information Search and Retrieval}{Digital Libraries} \vspace{-0.4cm} 


\keywords{Scientific impact; Science of science; Citation prediction; Popularity prediction}
}

\section{Introduction}
\label{sec:intro}

Integral to the success of scientific research is the publication and dissemination of impactful work and findings.

Every scientific researcher leaves his or her own indelible mark on an ever-expanding body of literature through a personal track-record of academic publications. The impact of each of these publications---both to a field of research and, by extension, to the reputation of the author---can be influenced by a variety of factors. For example, significant research work may start from a small number of exploratory papers that builds up to pioneering work within a field, resulting in a series of less impactful papers that serve as stepping-stones to those of greater impact. Or a researcher may publish in different fields each with differing audiences and levels of popularity, resulting in publications on some topics receiving more attention than those on others. Or, along with many publications that incrementally advance a field, a researcher may produce a groundbreaking work that transforms the field or even stimulates a new research area. As a result of such factors, a researcher's body of work is likely to be comprised of publications of varying impact. Accordingly, the impact of any particular publication can be difficult to predict.

Yet while the impact of individual publications may vary, a researcher's influence and productivity is measured by his or her body of work as a whole. Often, the researcher's total number of citations is used as a measure of impact, while the researcher's total number of publications is used as a measure of productivity. However, while these simple measures can be useful, they can also have drawbacks. For example, the total number of citations can be skewed by the impact of a solitary well-cited, impactful paper. Similarly, the total number of publications can be increased by a large number of poorly cited papers. Moreover, as citations demonstrate a power law distribution, with the vast majority of publications receiving few citations, these simple measures are difficult to estimate using traditional regression analysis \cite{Radicchi:PNAS08,Cheng:WWW14}. Thus, answering the question of how many citations a given paper will receive is often ineffective for practical purposes.

In light of these difficulties and limitations, we instead address an analogous question asked by many academic researchers in the course of writing a paper: ``Will \textit{this} paper increase my \hindex?''

The \hindex\ is an index that attempts to measure both the productivity and impact of the published work of a scientist or scholar. The index was suggested in 2005 by Jorge E. Hirsch as a tool for determining theoretical physicists' relative quality, and is based on the distribution of citations received by a given researcher's publications. As described by Hirsch: ``A scientist has index $h$ if $h$ of his/her $N_p$ papers have at least $h$ citations each, and the other $(N_p - h)$ papers have no more than $h$ citations each'' \cite{Hirsch:05}. In other words, a scholar with an index of $h$ has published $h$ papers, each of which has been cited in other papers at least $h$ times. Thus, the \hindex\ is a function of both the number of publications and the number of citations per publication.

The index is designed to improve upon simpler measures such as the total number of citations or publications. Moreover, because only the most highly cited articles contribute to the \hindex, its determination is a relatively simpler process. Hirsch has demonstrated that $h$ has high predictive value for whether a scientist has won honors like National Academy membership or the Nobel Prize \cite{Hirsch:05}. As a result of its simplicity and predictive value, the \hindex\ has become a \textit{de facto} standard for measuring academic performance and scientific impact.


\vpara{Contributions.} In this work, we formalize the question of whether a paper will influence an author's \hindex\ as a novel scientific impact prediction problem. Our prediction task is to determine whether a given paper will, within a pre-defined timeframe, increase the \hindex\ of its primary author (i.e., the researcher with the maximum \hindex\ among the paper's author list). Factors such as the researcher's current influence, the publication topic, and the publication venue may, among many other factors, play a role in determining the degree to which the publication contributes to the researcher's influence. A resulting challenge is the interplay of such factors, which can confound attempts to generate effective predictions. Considerations such as the variability of the \hindex\ according to the ``academic age'' of a researcher, the widely differing citation conventions among different fields, and the co-authorship of researchers with differing \hindices\ can make it difficult to isolate the nature and degree to which a given researcher's \hindex\ is influenced by any particular factor. 
Our work focuses on addressing and overcoming these issues to generate novel, effective scientific impact predictions and to investigate precisely what role a variety of factors play in these predictions.

\begin{figure}[t]
\centering
\includegraphics[width=3in]{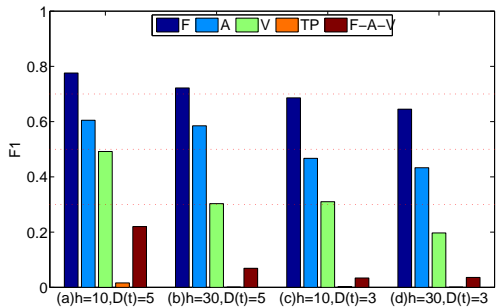}
\caption{\label{fig:intro-f1}
{\bf Predictability of whether one paper will increase the \hindex\ ($\geq$ $h$) of its primary author  within $D(t)$ years.} 
F: full factor set; A: only authority factors; V: only venue factors; TP: only topic popularity factors; F-A-V: full factors without authority and venue factors. 
}
\end{figure}

Given the task of predicting whether a publication will contribute to an author's \hindex, we find surprisingly strong performance for the problem of scientific impact prediction. Our results demonstrate that we can predict whether a paper will contribute to an author's \hindex\ within five years with an \fonescore\ of 0.776 as shown in Figure~\ref{fig:intro-f1}. Our study further finds that the most telling factors for determining whether a given paper will contribute to the primary author's \hindex\ are the author's authority on the publication topic and the venue in which the paper is published (see Figure~\ref{fig:intro-f1} blue bars). In contrast, we find that the popularity of the publication topic and the co-authors' \hindices\ are surprisingly inconsequential for determining whether the paper will contribute to the primary author's \hindex\ (see Figure~\ref{fig:intro-f1} orange and red bars). We also find that (1) the contribution of papers to researchers with higher \hindices\ is more difficult to predict than it is for researcher's with lower \hindices\ (cf. Figure~\ref{fig:intro-f1} a vs. b and c vs. d); and (2) the task is more predictable given a long timeframe than a short one (cf. Figure~\ref{fig:intro-f1} a vs. c and b vs. d). 
Overall, our findings unveil mechanisms for quantifying scientific impact and provide concrete suggestions to researchers for better expanding their scientific influence and, ultimately, for more effectively ``standing on the shoulders of giants.''

A potential concern with this work is that by targeting the \hindex, our findings may result in unintended side effects by a principle referred to by economists as Goodhart's Law. The law essentially says that once a measure is chosen as a performance indicator it begins to lose value, as measurement can distort the practice being measured. Restated more succinctly by \cite{strathern1997improving}, it means that ``when a measure becomes a target, it ceases to be a good measure.'' Be that as it may, we believe that understanding the characteristics of scientific impact measures is imperative to their informed use, and \textit{in no way should our research be construed as advocating the use of the \hindex\ as a deciding factor in one's research pursuits}.


\section{Data and Problem}
\label{sec:problem}

In this section, we first describe the academic data we use, then characterize the \hindex\ from the data, and finally formalize our predictive task---that is, scientific impact prediction.

\subsection{Data Description}

In this paper, we use the real-world academic dataset\footnote{\small The dataset is publicly available at \href{http://arnetminer.org/citation}{ArnetMiner Citation} and \href{http://arnetminer.org/AMinerNetwork}{ArnetMiner APC}.} from ArnetMiner \cite{Tang:08KDD}, which is the world-leading free online service for academic social network analysis and mining.
The dataset contains 1,712,433 authors with 2,092,356 papers from computer science venues held between 1960 and 2012.
Each paper includes information on the title, abstract, authorship, references, and publication venue and year.
In total, the dataset represents 4,258,615 collaboration relationships and 8,024,869 citation relationships.

\begin{figure}[t]
\centering
\subfigure[\scriptsize Distribution of citations counts]{
\hspace{-0.2in}
\label{figsub:citation-powerlaw}
\includegraphics[width=1.75in]{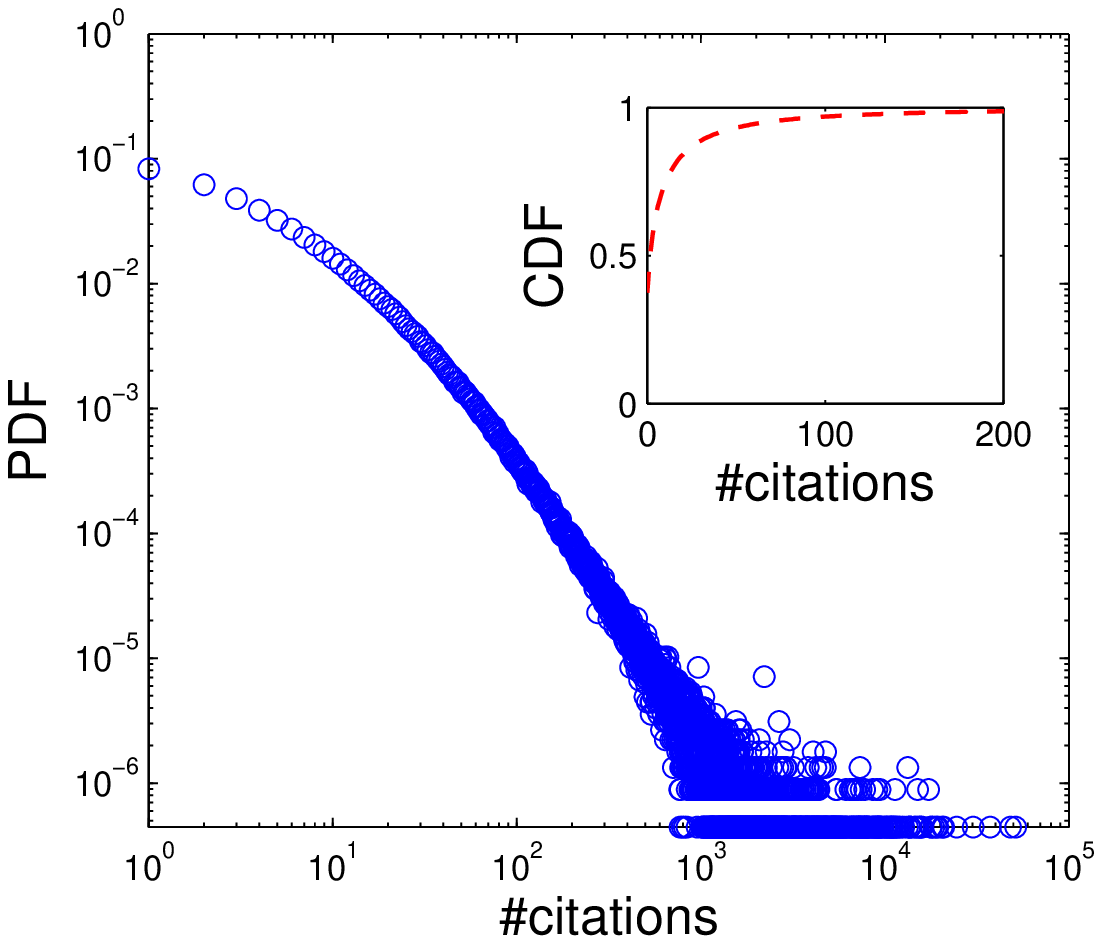}
}
\hspace{-0.1in}
\subfigure[\scriptsize \hindex\ distribution]{
\label{figsub:hindex-powerlaw}
\includegraphics[width=1.75in]{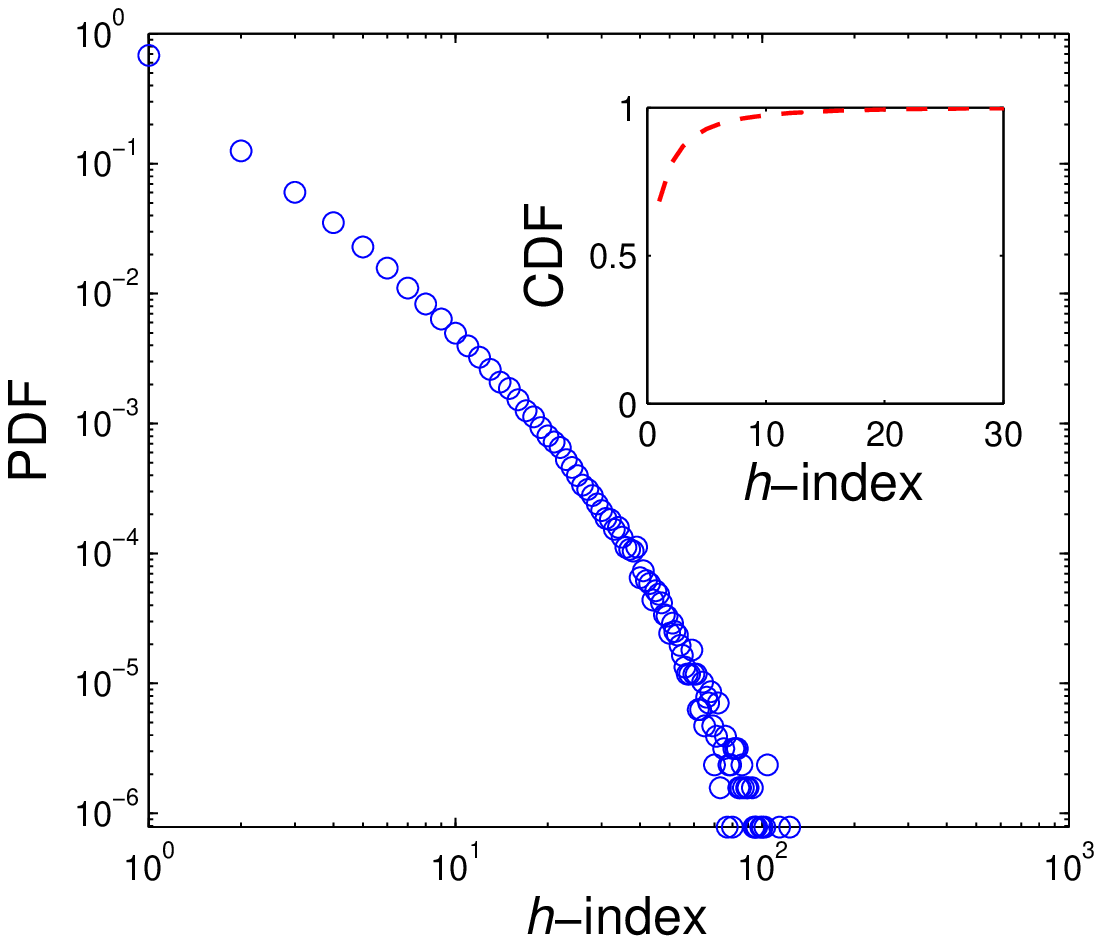}
\hspace{-0.2in}
}
\caption{\label{fig:powerlaw}
{\bf Distributions of the citation counts of papers and the \hindices\ of authors.} The heavy-tailed distributions indicate that the prediction of both metrics is not fit for regression problems.
In this dataset, 6.91\% (154,985) of the papers obtain more than 50 citations and 0.0125\% (159) of the researchers have \hindices\ greater than 60.
}
\vspace{-0.2cm}
\end{figure}

Next, we briefly explore and report the data characteristics of the author-paper-citation data.
Figure~\ref{fig:powerlaw} shows the distributions of the number of citations for each paper and the \hindex\ of each author.
In our dataset, both metrics follow heavy-tailed distributions (i.e., distributions with a ``tail'' that is ``heavier'' than that of an exponential).
Moreover, only 6.91\% (154,985) of the papers have more than 50 citations, while 0.0125\% (159) of the researchers have an \hindex\ over 60.

\begin{figure}[t]
\centering
\subfigure[\scriptsize \hindex\ vs. $\#$papers]{
\hspace{-0.2in}
\label{figsub:h-numpapers}
\includegraphics[width=1.75in]{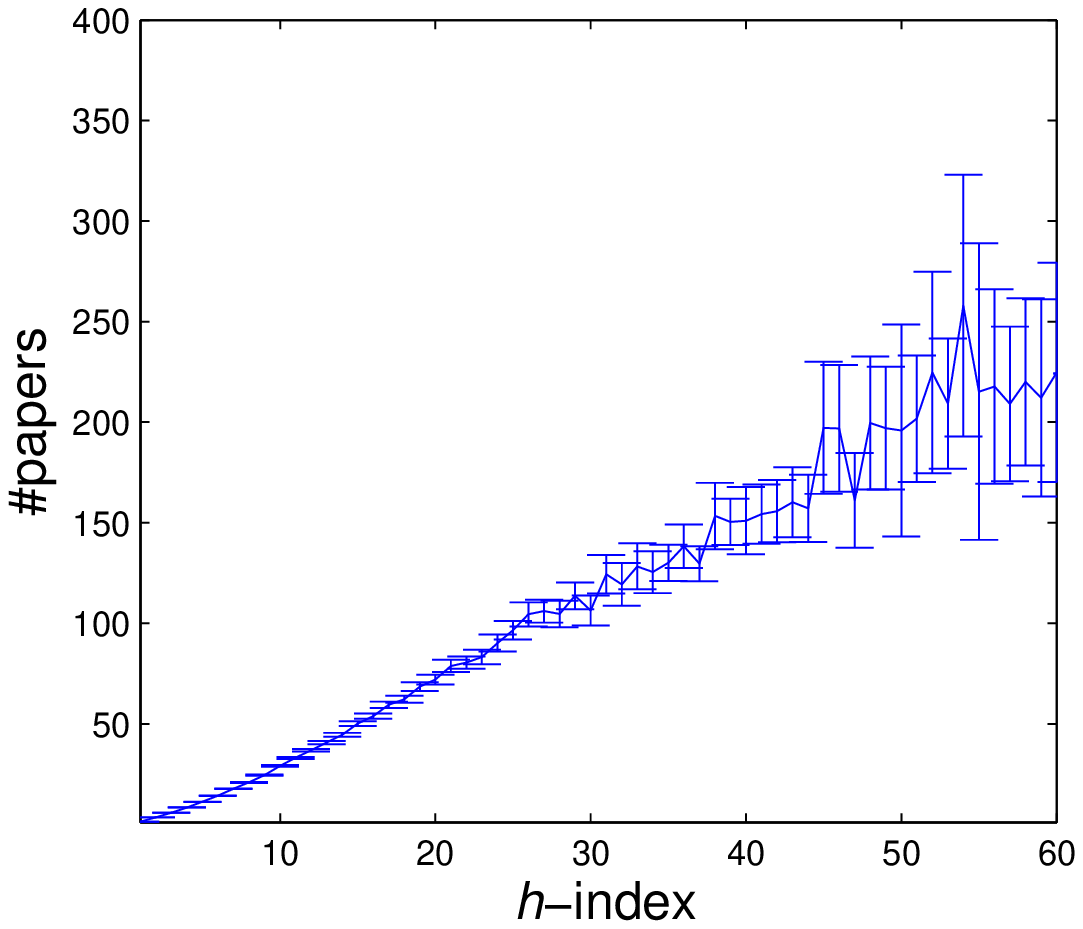}
}
\hspace{-0.1in}
\subfigure[\scriptsize \hindex\ vs. $\#$average citations]{
\label{figsub:h-avec}
\includegraphics[width=1.75in]{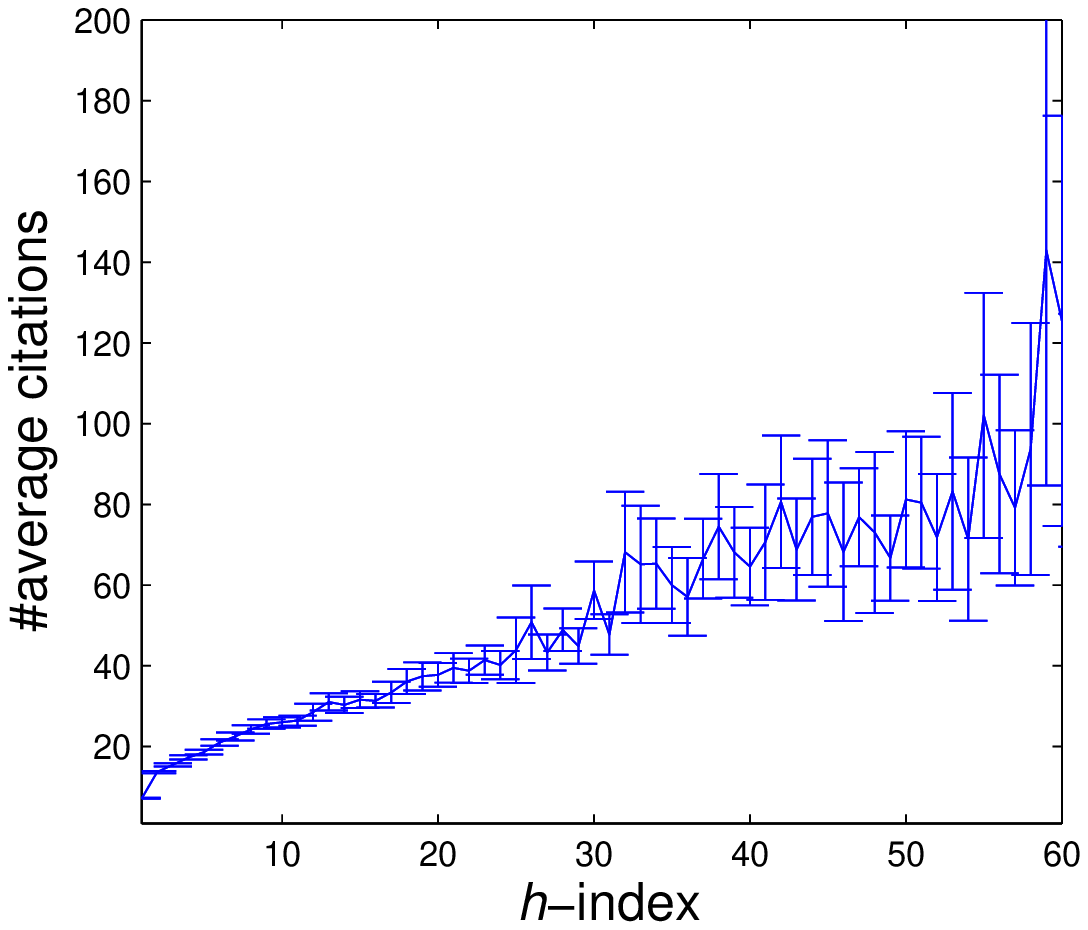}
\hspace{-0.2in}
}
\subfigure[\scriptsize \hindex\ vs. $\#$co-authors]{
\hspace{-0.2in}
\label{figsub:h-numco-authors}
\includegraphics[width=1.75in]{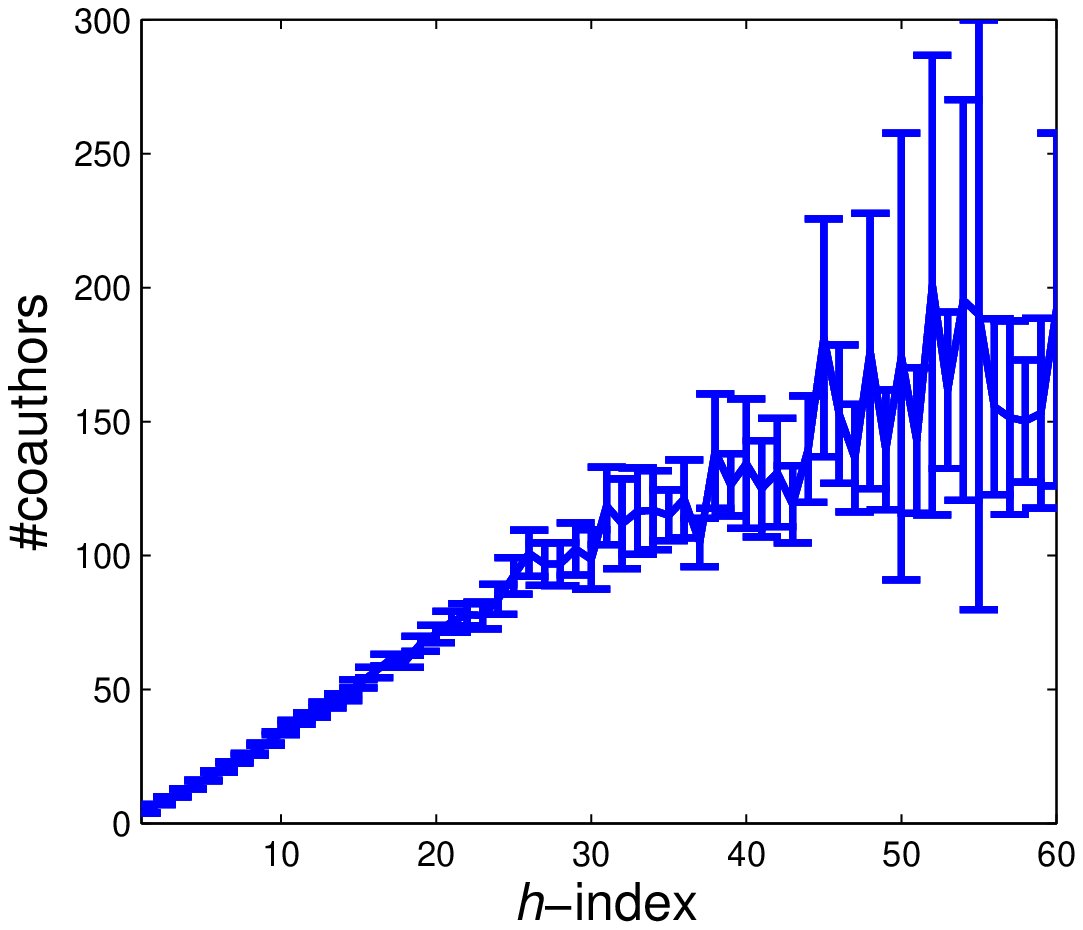}
}
\hspace{-0.1in}
\subfigure[\scriptsize \hindex\ vs. \hindex/$\#$papers]{
\label{figsub:h-hratio}
\includegraphics[width=1.75in]{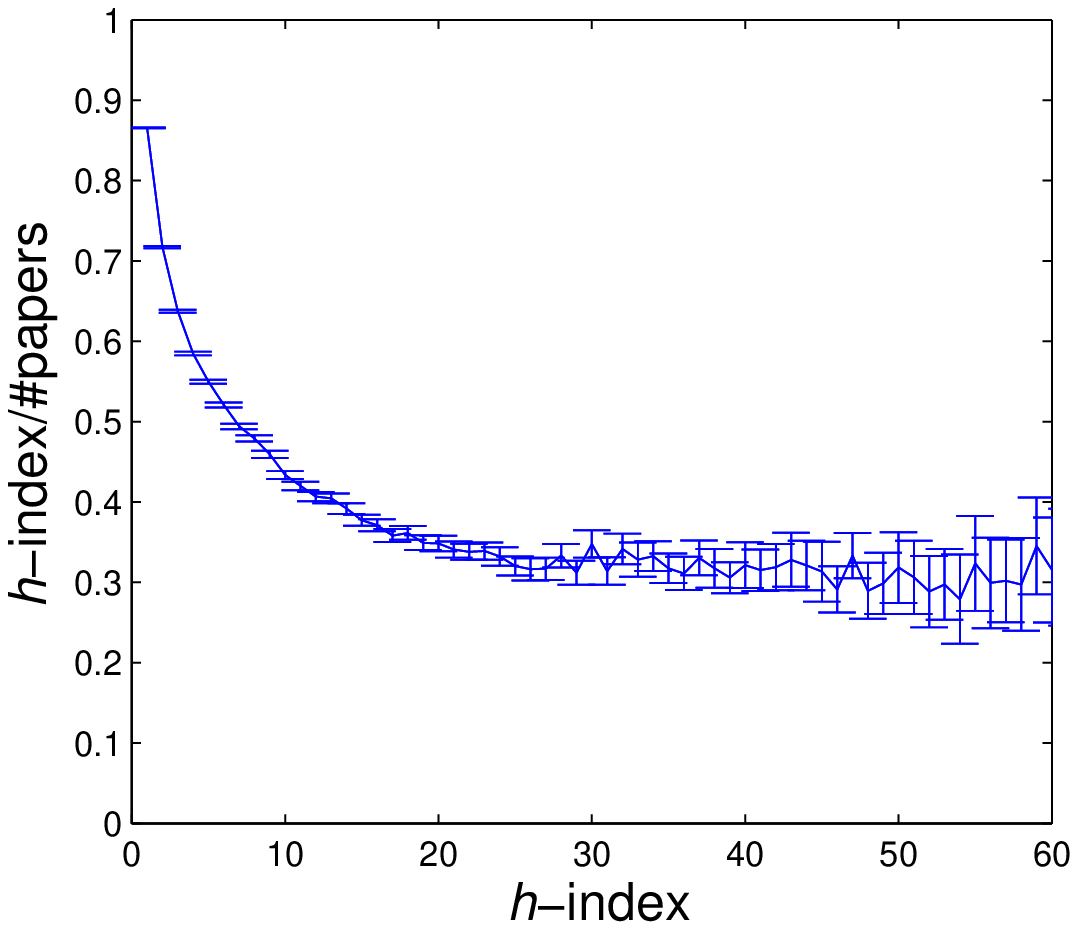}
\hspace{-0.2in}
}
\subfigure[\scriptsize \hindex\ vs. co-authors' \hindex]{
\hspace{-0.2in}
\label{figsub:h-hco-author}
\includegraphics[width=1.75in]{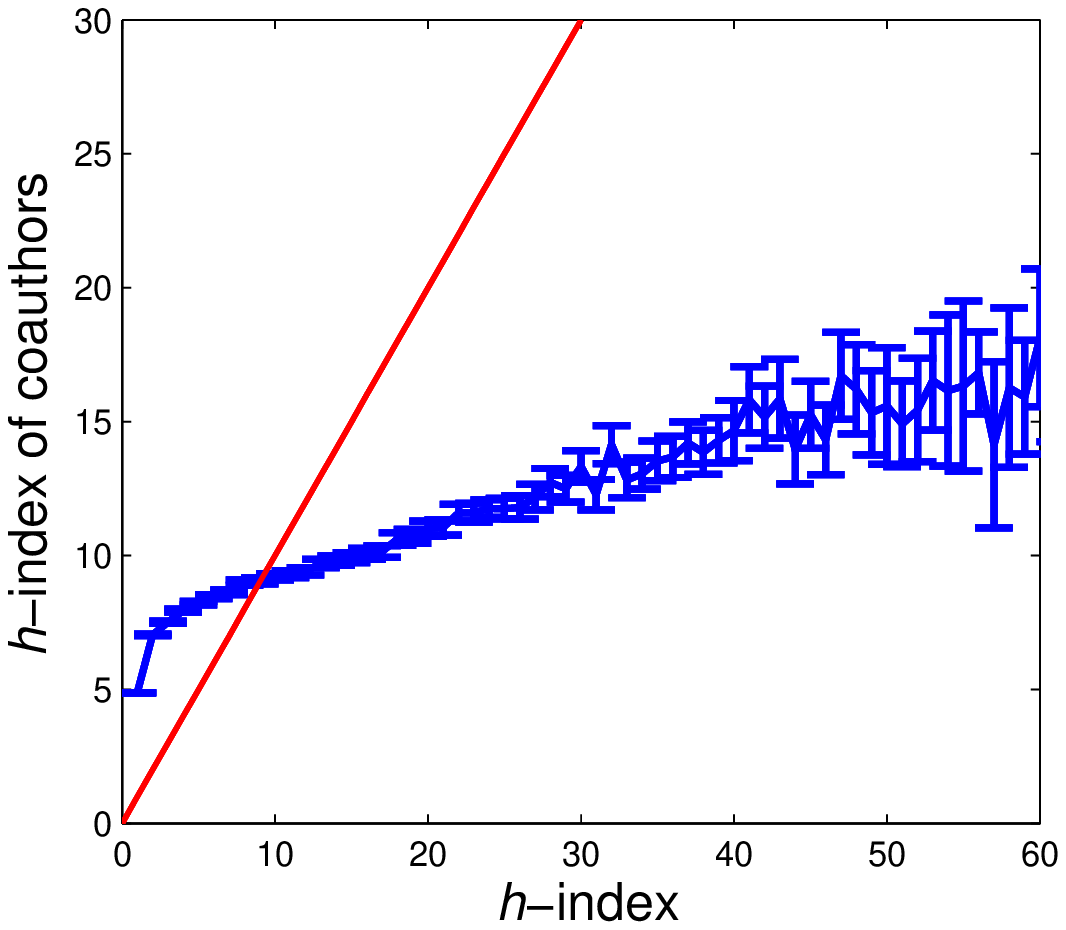}
}
\hspace{-0.1in}
\subfigure[\scriptsize \hindex\ vs. $\#$years]{
\label{figsub:h-years}
\includegraphics[width=1.75in]{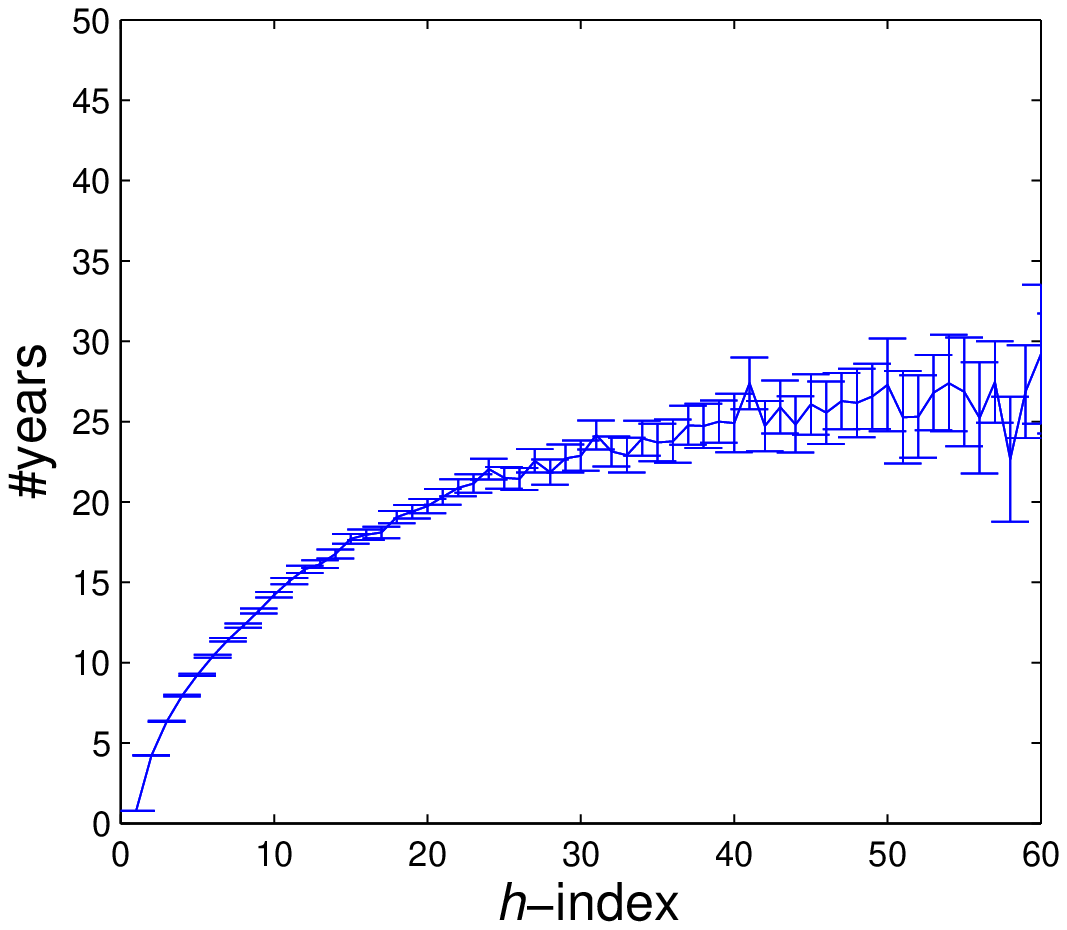}
\hspace{-0.2in}
}
\caption{\label{fig:citation-h}
{\bf Characterizations of the \hindex. }
(b) The average number of citations for each author is larger than her/his \hindex. 
(d) The ratio between one's \hindex\ ($\geq$ 20) and her/his number of papers stabilize at 0.3. 
(e) Typically, the author's \hindex\ becomes larger than the co-authors' \hindices\ at the expected point of the author's Ph.D. graduation. 
(f) The rate at which the \hindex\ increases itself increases as the length of time spent in academia becomes longer (\textit{i.e., the rich get richer}).
}
\vspace{-0.2cm}
\end{figure}

\vpara{Characterizations of the \hindex.}
The \hindex\ is defined as the number of papers with citation number $\geq h$, which is a useful index to characterize the scientific output of a researcher~\cite{Hirsch:05}.
Figure~\ref{fig:citation-h} presents the basic characteristics of scientific impact in terms of \hindex.
Positive linear relationships are clearly observed between the \hindex\ and the number of papers, average number of citations, and number of co-authors in Figures~\ref{figsub:h-numpapers}, \ref{figsub:h-avec}, and \ref{figsub:h-numco-authors}, respectively.
Figure~\ref{figsub:h-avec} shows that the average number of citations for each author is larger than her/his \hindex.
Figure~\ref{figsub:h-hratio} illustrates the ratio between one's \hindex\ and the number of papers conditioned on her/his \hindex.
This ratio initially decreases, stabilizing after an \hindex\ of about 20.
From Figure~\ref{figsub:h-hco-author}, we see that the \hindices\ of one's co-authors increase as her/his \hindex\ grows.
The reversion point between author and co-author \hindices\ occurs when one's \hindex\ reaches about 8 or 9, a typical point at which Ph.D. students graduate.
Finally, in Figure~\ref{figsub:h-years}, we examine the interplay between authors' \hindex\ and the length of time they spend in academia (the date difference between one's first and last publications).
Clearly, we see that the increase of \hindex\ is initially slow upon first entering academia.
As the \hindex\ increases, the accumulations of influence, resources, connections, and previous papers further drive one's \hindex\, and scientific impact expands at an increasingly rapid rate.
In other words, the aphorism that ``the rich get richer'' is demonstrated in academia, whereby the influence of individuals who have already accumulated a great deal of influence increases at a disproportionally quick rate.
All characteristics are observed at a 95\% confidence interval.

\subsection{Problem Definition}

Traditionally, the task of scientific impact prediction is formulated as a regression problem for predicting citation counts \cite{Yan:JCDL2012}.
However, the intrinsically heavy tailed distribution of citation counts, demonstrated in Figure~\ref{figsub:citation-powerlaw}, make such predictions necessarily skewed \cite{Cheng:WWW14}.
Alternatively, given a threshold $c$ for the citation count, the task can be formulated as a classification problem---that is, a problem of inferring whether the number of citations for a given paper will reach $c$.
Even with this formulation, the way in which citation counts are distributed presents inherent challenges, including how to determine the threshold $c$ and how to avoid the overrepresentation or underrepresentation of different papers.
Similar issues also arise for predicting the \hindex\ of each author.

Our approach is inspired by the work of \cite{Cheng:WWW14}, where the authors consider the problem of Facebook cascade growth prediction. They formulate the problem as a binary classification task where, given the first $k$ reshares of content within a cascade, they are tasked with predicting whether the cascade will reach $2k$. For our work, we formulate an analogous task:  Given a paper, we predict whether that paper will increase the \hindex\ of the authors in the future. We formalize the problem of scientific impact prediction next.

\begin{problem}
\textbf{Scientific Impact Prediction:} Given the publication corpus $C$ before timestamp $t$, each paper (document) $d \in C$ published in timestamp $t$, and the primary author's \hindex\ (max-\hindex) among the authors of $d$, the task is to predict whether the number of citations for each paper $d$ will reach max-\hindex\ after a given time period $\Delta t$.
\end{problem}

The primary author is defined as the author of the given paper with the highest \hindex.
We address the issues noted above by using a local threshold max-\hindex\ for each paper's future citation count.
For example, according to Google Scholar, as of Nov. 25th, 2014, Dr. Yizhou Sun and Dr. Jiawei Han's \hindices\ are 19 and 125, respectively.
By setting $t=2014$ and $\Delta t=3$ years, we aim to predict whether each of Dr. Sun's papers (assuming her \hindex\ is larger than each of her co-authors') published in 2014 will get more than 19 citations by 2017.
But if one paper is co-authored by Dr. Sun and Dr. Han, the task becomes to infer whether this paper's citation count will reach 125 by 2017.
Realistically, the authors' \hindices\ may increase during the duration $\Delta t$, but it is intractable to capture this evolution in detail.
In this sense, we consider the objective of collecting citations for each paper as static and leave the evolving case for future work.

This proposed problem of scientific impact prediction is completely different from the cascade growth prediction problem \cite{Cheng:WWW14}, which needs the observation of the first $k$ reshares (citations), and further predicts the future reshare counts.
Our problem is also fundamentally different from the traditional citation count prediction problem \cite{Yan:JCDL2012}, which focuses on the regression task to predict scientific impact.
In contrast, our problem is to predict each paper's future impact conditioned on the authors. The chief advantage of this proposed formulation is that it can be leveraged for a variety of real-world applications, such as author \hindex\ and popularity prediction~\cite{Shen:AAAI14}, expert finding~\cite{Zhang:07DASFAA}, and credit allocation~\cite{Shen:PNAS14,Kleinberg:STOC11}. \\

\section{Scientific Impact Factors}
\label{sec:factor}

\begin{table*}[t]
\caption{\bf{Factor Definition.}
We employ six categories of factors, comprised of author, topic, reference, social, venue, and temporal attributes. 
max-\hindex\ denotes the \hindex\ of the primary author (i.e., the author with the maximum \hindex) of a given paper.
}
\label{tb:factors}
\centering
\renewcommand\arraystretch{1.3}
\begin{tabular}{@{}l|l|l}

\hline

& Factor & Description \\ \hline
\multirow{7}*{Author} & \textit{A-first-max} & The first author's \hindex\ divided by the max-\hindex. \\
& \textit{A-ave-max}     & The average \hindex\ of all authors divided by the max-\hindex. \\
&\textit{A-sum-max}      & The sum of \hindices\ divided by the max-\hindex. \\
& \textit{A-first-ratio} & The ratio between max-\hindex\ and the number of papers attributed to the first author. \\
&\textit{A-max-ratio}    & The ratio between max-\hindex\ and the number of papers attributed to the primary author. \\
&\textit{A-num-authors}  & The number of authors of the given paper. \\
&\textit{A-num-first}    & The number of papers by the first author. \\ 

\hline

\multirow{7}*{Content} & \textit{C-popularity} & The average number of citations over different topics (see Eq.~\ref{eq:popularity}). \\
&\textit{C-popularity-ratio} & The average number of citations over different topics divided by the max-\hindex. \\
&\textit{C-novelty}          & The topic novelty of this paper (see Eq.~\ref{eq:novelty}). \\
&\textit{C-diversity}        & The topic diversity of this paper (see Eq.~\ref{eq:diversity}). \\
&\textit{C-authority-first}  & The consistence between the first author's authority and this paper (see Eq.~\ref{eq:authority}). \\
&\textit{C-authority-max}    & The consistence between the primary author's authority and this paper. \\
&\textit{C-authority-ave}    & The average consistence between each author's authority and this paper. \\

\hline

\multirow{2}*{Venue} & \textit{V-ratio-max} & The ratio between the number of papers $\geq$max-\hindex\ citations divided by the max-\hindex. \\ 
&\textit{V-citation} & The average number of citations of all references divided by the max-\hindex. \\

\hline

\multirow{4}*{Social}  & \textit{S-degree} & The number of co-authors of the paper's authors. \\
&\textit{S-pagerank}   & The PageRank values of the paper's authors in the weighted collaboration network. \\
&\textit{S-h-co-author} & The average \hindex\ of co-authors of the paper's authors divided by the max-\hindex. \\
&\textit{S-h-weight}   & The weighted average \hindex\ of co-authors of the paper's authors divided by the max-\hindex. \\

\hline

\multirow{2}*{Reference} & \textit{R-ratio-max} & The ratio between the number of references $\geq$max-\hindex\ and the total number of references. \\
&\textit{R-citation} & The average number of citations divided by the maximum \hindex. \\

\hline

\multirow{4}*{Temporal} & \textit{T-ave-h} & The average $\Delta$\hindices\ of the authors between now and three years ago. \\ 
&\textit{T-max-h}   & The maximum $\Delta$\hindex\ between now and three years ago. \\
&\textit{T-h-first} & The $\Delta$\hindex\ of the first author between now and three years ago. \\
&\textit{T-h-max}   & The $\Delta$\hindex\ of the max-\hindex\ author between now and three years ago. \\

\hline

\end{tabular}
\end{table*}

\begin{figure*}[t]
\centering
\subfigure[\scriptsize Author factors]{
\hspace{-0.3in}
\label{figsub:cc-h-author}
\includegraphics[width=1.75in]{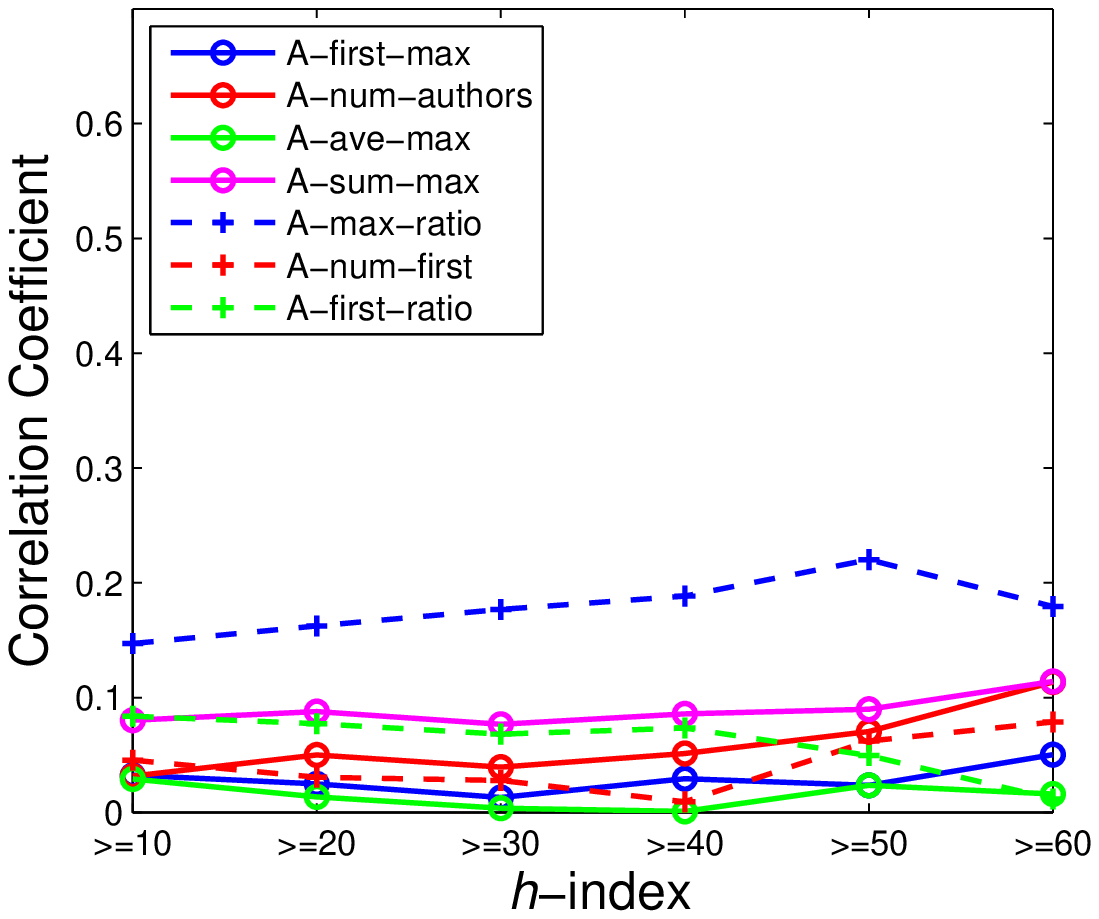}
}
\hspace{-0.1in}
\subfigure[\scriptsize Content factors]{
\label{figsub:cc-h-content}
\includegraphics[width=1.75in]{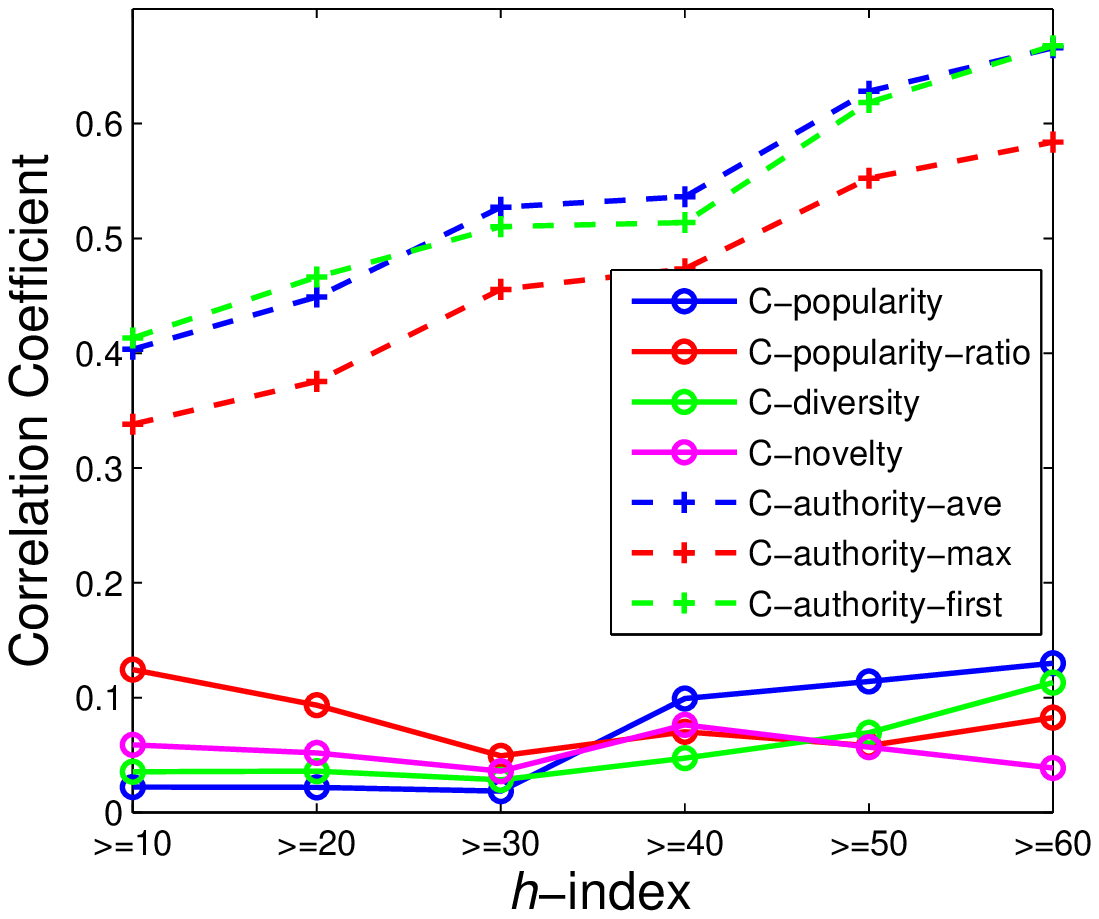}
}
\hspace{-0.1in}
\subfigure[\scriptsize Social and venue factors]{
\label{figsub:cc-h-social-venue}
\includegraphics[width=1.75in]{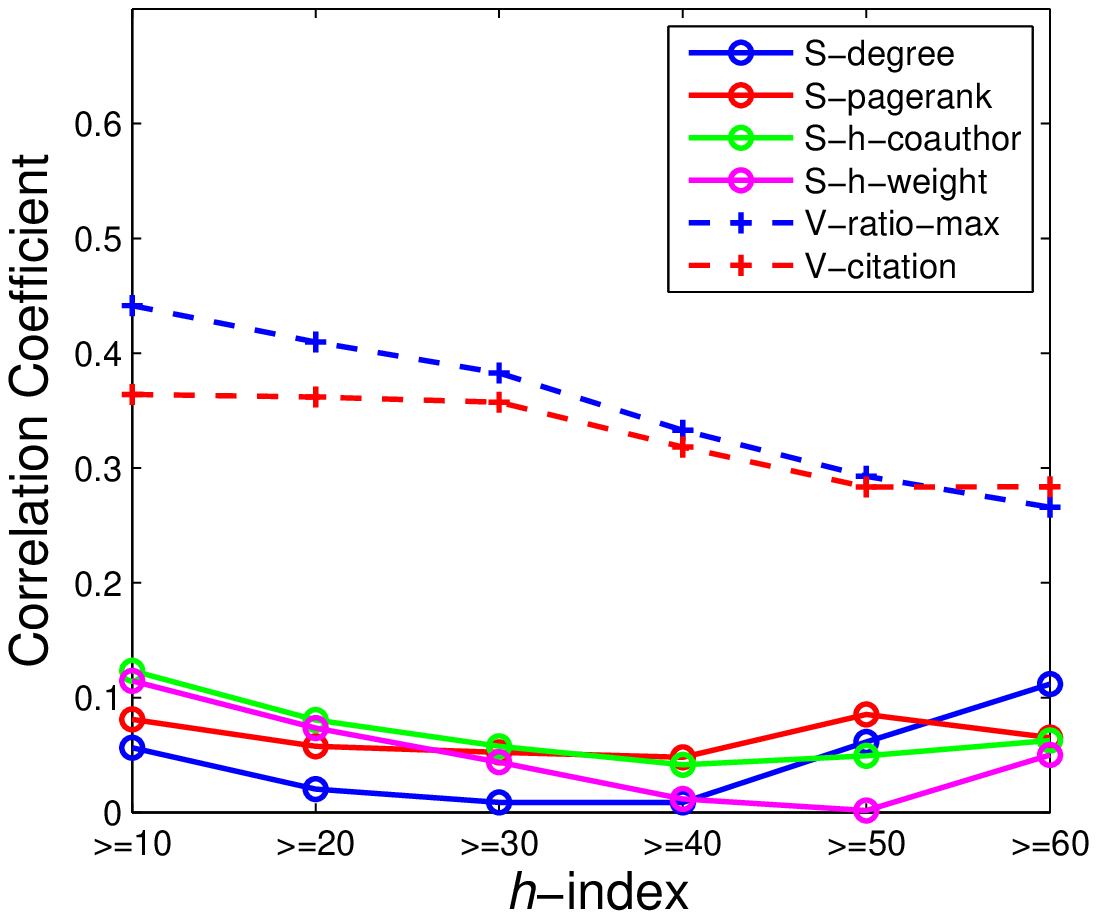}
}
\hspace{-0.1in}
\subfigure[\scriptsize Reference and temporal factors]{
\label{figsub:cc-h-ref-temp}
\includegraphics[width=1.75in]{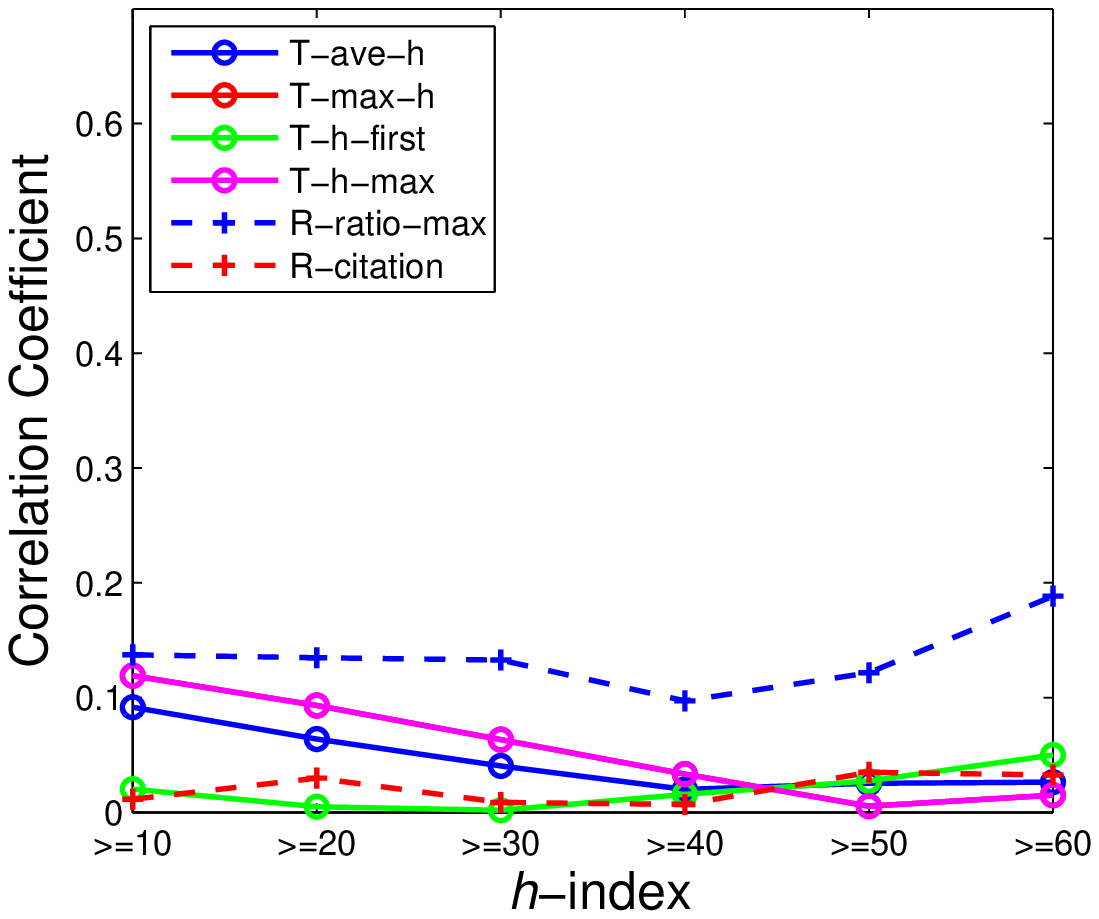}
\hspace{-0.2in}
}
\caption{\label{fig:cc-h}
{\bf Factor correlation analysis when predicting for scholars with different \hindices.}
$t$=2007 and $\Delta t$=5 years. 
Author's authority on a subject and published venue are the most highly correlated factors.
}
\end{figure*}

\begin{figure*}[t]
\centering
\subfigure[\scriptsize Author factors]{
\hspace{-0.3in}
\label{figsub:cc-year-author}
\includegraphics[width=1.75in]{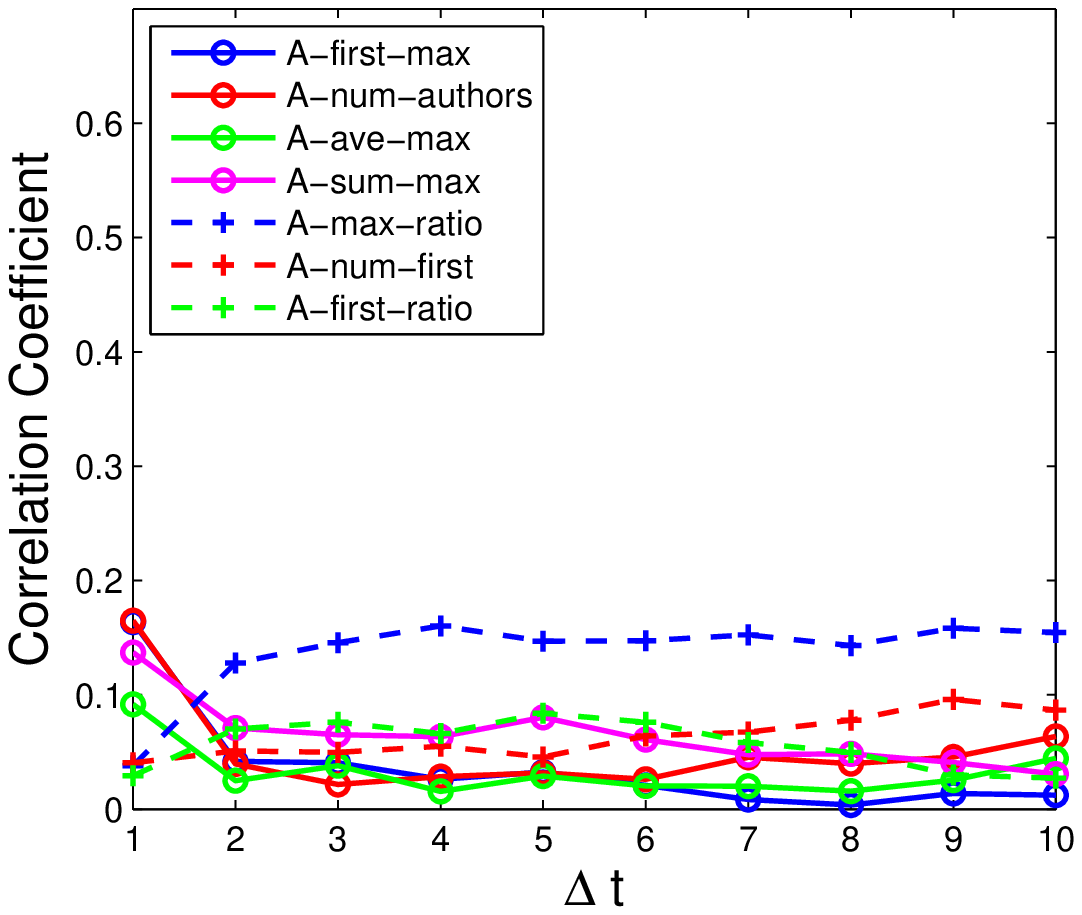}
}
\hspace{-0.1in}
\subfigure[\scriptsize Content factors]{
\label{figsub:cc-year-content}
\includegraphics[width=1.75in]{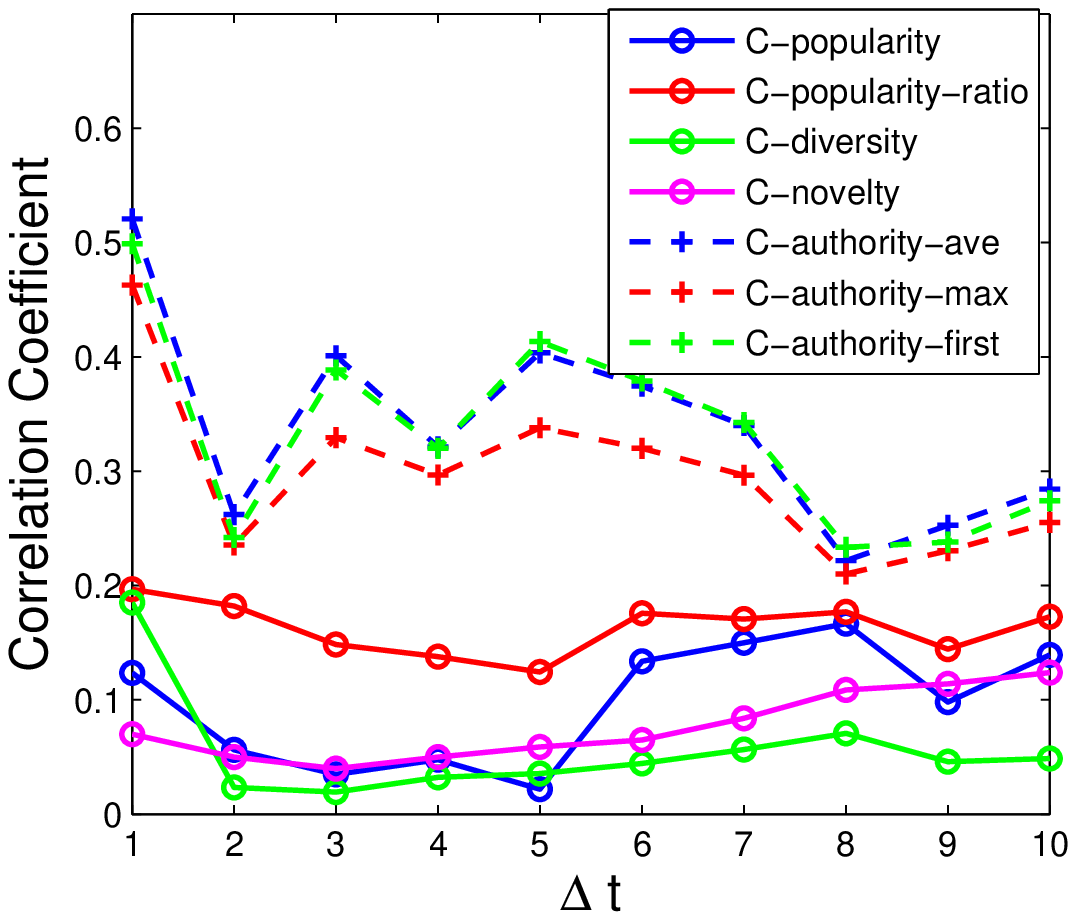}
}
\hspace{-0.1in}
\subfigure[\scriptsize Social and venue factors]{
\label{figsub:cc-year-social-venue}
\includegraphics[width=1.75in]{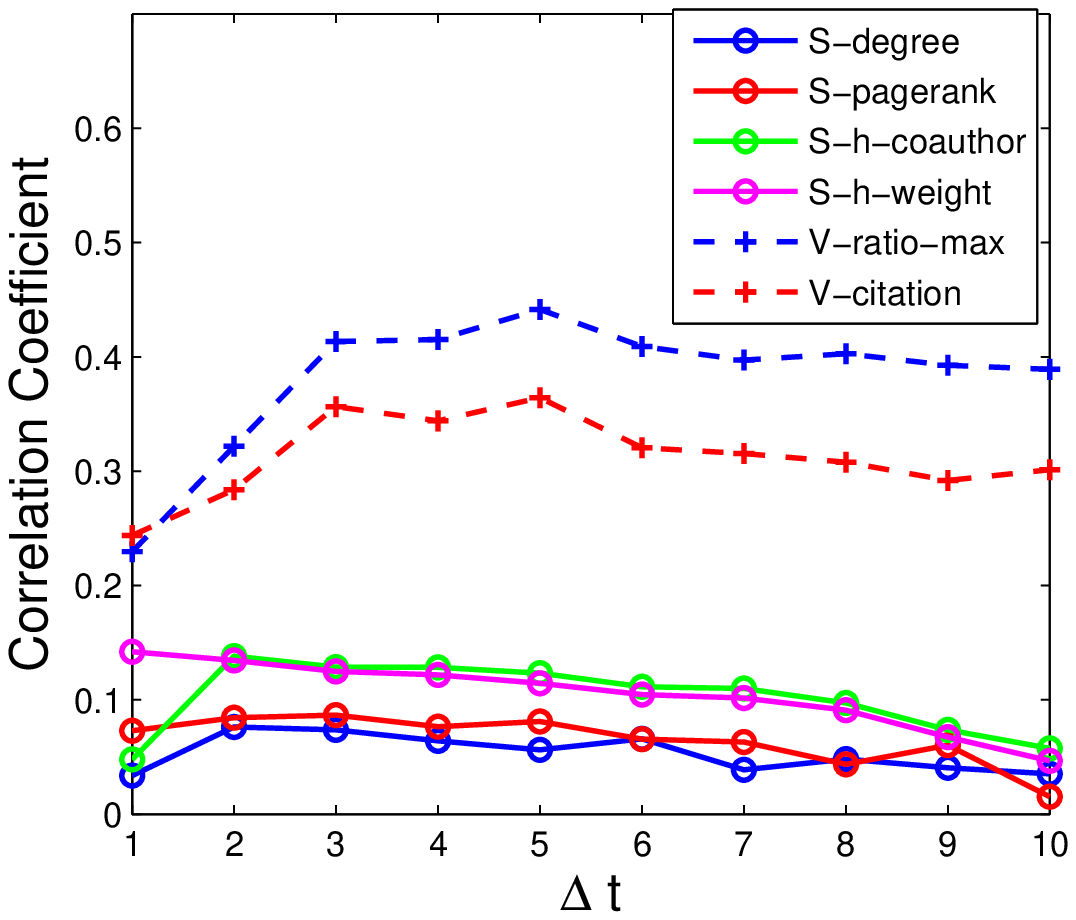}
}
\hspace{-0.1in}
\subfigure[\scriptsize Reference and temporal factors]{
\label{figsub:cc-year-ref-temp}
\includegraphics[width=1.75in]{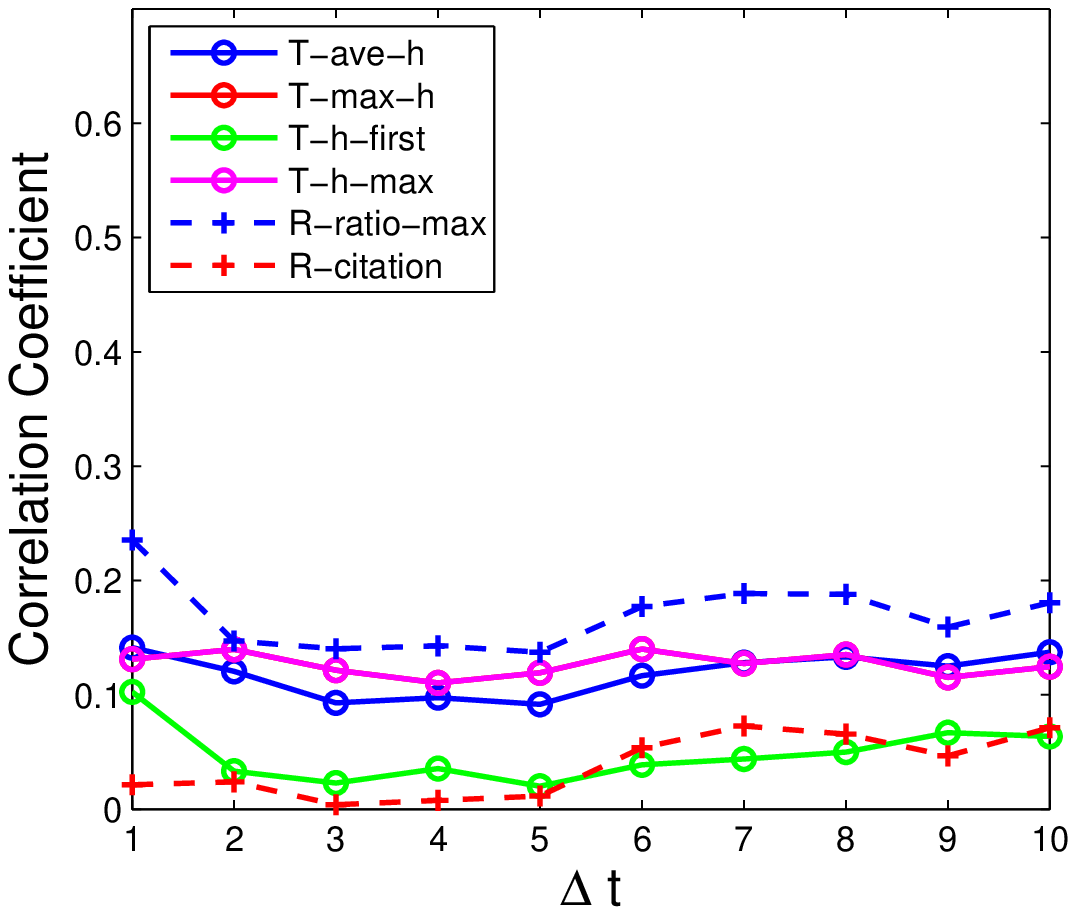}
\hspace{-0.2in}
}

\caption{\label{fig:cc-year}
{\bf Factor correlation analysis as the time period $\Delta t$ is varied.} 
Note that $t$=2007 and a minimum threshold for the primary author's \hindex\ is set to 10.
}
\end{figure*}

To quantify scientific impact, it is natural to use the number of citations obtained by each paper and its authors. 
Recall that given a paper $d$, our objective is to predict whether the number of citations $c_{d}$ obtained within a given time period $\Delta t$ will be larger than its primary author's \hindex\ (denoted as max-\hindex). 
We investigate the factors that drive a paper's citation count to become greater than its primary author's \hindex, including the paper's author(s), content, published venue, and references, as well as social and temporal effects related to its author(s).
Table \ref{tb:factors} lists the six different groups of factors investigated.
Figures~\ref{fig:cc-h} and \ref{fig:cc-year} show the importance of different factors as evaluated by correlation coefficients. 
In Figure~\ref{fig:cc-h}, we present the changes of factor importance as predicted for scholars with different \hindices.
In Figure~\ref{fig:cc-year}, the changes of factor correlation are plotted as the time period $\Delta t$ is varied.

\subsection{Author Factors}
The predictive task for each paper naturally depends on the authors themselves, including both the primary author and his or her co-authors. 
Prior work has been devoted to examining the interplay between scientific impact (number of citations) and the average values of authors' attributes \cite{Yan:JCDL2012,Castillo:2007}. 
Given our problem formulation, we also investigate the attributes of the primary author of each paper, such as the ratio of the author's previous papers contributing to his or her \hindex.
Additionally, as the first author usually leads the collaboration, the spread of a paper's scientific impact also relies highly on the impact of the first author. 
We thus consider the probability that the number of citations obtained by the first author's previous publications is greater than the primary author's \hindex.
Further, the overall impact of all co-authors should have the potential to influence a paper's quality and popularity.
Thus the sum of all authors' \hindices\ is used to simulate their overall impact.
Since each paper has a unique local threshold (max-\hindex) in our classification problem, we calculate the relative value of each factor by dividing it by the max-\hindex.
For example, the first author's \hindex\ divided by the max-\hindex\ is designed to model the author's effect on the success of the paper.
Moreover, the author's productivity, i.e., the number of his or her previous publications, has a positive effect on the future citation counts of the paper due to self-citation behavior~\cite{Bethard:CIKM2010}.

\vpara{Significance.}
Figure~\ref{figsub:cc-h-author} and \ref{figsub:cc-year-author} clearly show that \textit{A-max-ratio} of one paper---i.e., the ratio between the max-\hindex\ and the number of papers attributed to its primary author---has the highest correlation with the probability that it will collect at least max-\hindex\ citations among all seven author factors. 
We also observe that the author factor correlations are not sensitive to the two parameters, illustrated by the lower thresholds for the primary author's \hindex\ and $\Delta t$.

\subsection{Content Factors}
Aside from the attributes of its authors, another intuitive factor affecting a paper's success is its content. 
Topic modeling is a widely used method for extracting and mining the content of literature and can be used to extract ``topics'' that occur in a collection of documents. One of the most popular topic modeling methods is known as Latent Dirichlet Allocation (LDA), a generative probabilistic approach that views each document as a mixture of various topics \cite{Blei:03}.
Similar to previous work on modeling citation counts~\cite{Yan:JCDL2012}, we run a 100-topic LDA model on the title and abstract of the corpus $C$ published before time $t$ and the target papers published at time $t$, which returns the probability distribution $p(z|d)$ over topics $z\in Z$ assigned for each paper $d$.
We define the following features based on each paper's topic distribution, including popularity, novelty, diversity, and authority.

First, we consider that as popular topics attract more attention, it is relatively easy for papers related to such topics to accrue citations. 
To capture this effect, we quantify the popularity of each topic $z$ across the overall corpus by
$$\small popularity(z) = \sum_{d \in C} p(z|d) \times c_d,$$
where $p(z|d)$ is the probability that paper $d$ distributes on topic $z$ and $c_d$ is the number of citations $d$ collects until the timestamp $t$.
Given one target paper $d_t$ at time $t$, its popularity is defined as
\begin{align}
\small
\textit{C-popularity}(d_t) = \sum_{z \in Z} popularity(z) \times p(z|d_t).
\label{eq:popularity}
\end{align}

Second, the novelty of a paper is considered an essential factor in assessing its contribution to the scientific community. 
Previous work assumes that the novelty of an article can be determined by measuring the difference between its content and that of its references~\cite{Yan:JCDL2012}.
We utilize the Kullback-Leibler divergence~\cite{Kullback:51} to capture the sum of the difference between $d_t$'s topic distribution and the topic distribution of each of its references. Specifically,
\begin{align}
\small
\textit{C-novelty}(d_t) = \frac{\sum_{d_r \in R} KL(p(Z|d_t), p(Z|d_r)) }{|R|} \label{eq:novelty}
\end{align}
\begin{align}
\small
KL(p(Z|d_t), p(Z|d_r)) = \sum_{z\in Z} \log \frac{p(z|d_t)}{p(z|d_r)} p(z|d_t) \nonumber,
\end{align}
where $R$ is the set of $d_t$'s references.

Third, the topic diversity of a paper, defined as the breadth of its topic distribution, is able to distinguish between different types of papers, such as between surveys and technical work.
We follow the definition of diversity in~\cite{Yan:JCDL2012} as calculated by Shannon entropy:
\begin{align}
\small
\textit{C-diversity}(d_t) = \sum_{z\in Z} -p(z|d_t)\log p(z|d_t).
\label{eq:diversity}
\end{align}

Fourth, Kleinberg has pointed out that, in a hyperlinked web environment, a good authority represents a page that is linked by many hubs~\cite{Kleinberg:99}. 
Similarly, authority in academia is denoted by being highly cited by others in a specific domain of expertise.
To measure the authority of researcher $a$ on topic $z$, we propose the following definition:
$$authority(a,z) = \sum_{d \in C_a} p(z|d) \times c_d,$$
where $C_a$ is the researcher $a$'s previous publications. 
Therefore, given the target paper $d_t$, the author's authority is distributed over the topic distribution of $d_t$, i.e.,
\begin{align}
\small
\textit{C-authority}(d_t, a) = \sum_{z\in Z} p(z|d_t) \times authority(a, z).
\label{eq:authority}
\end{align}
This definition of authority comes from the intuition that the correspondence between the paper's topic distribution and its authors' expertise help assure its quality.

\vpara{Significance.}
From Figures~\ref{figsub:cc-h-content} and \ref{figsub:cc-year-content}, we find that there is a strong correlation between a paper's probability of contributing to its primary author's \hindex\ and the authority-based content factors from that paper.
Perhaps surprisingly, the current topic popularity and the diversity or novelty of the paper's content have much lower correlations than its authors' expertise.
This observation suggests that, when authoring a paper, rather than following the current trend or hot topic, it is more important to instead focus on what one is good at.
Finally, the increase of the primary author's \hindex\ is accompanied by a corresponding rise in the importance of several authority factors, indicating that highly influential scholars are more easily recognized as the experts and authorities in their respective fields.

\subsection{Venue Factors}
Top venues attract high-quality submissions, and high-quality submissions elevate the reputation of their respective venues.
Google Scholar metrics show that different venues have large differences in their \textit{h5}-indices\footnote{\small \href{http://scholar.google.com/citations?view_op=top_venues&hl=en}{Google Scholar Metrics.} Accessed on Nov. 25th, 2014.}, which is defined as the \hindex\ when only considering articles published within the last 5 complete years.
For example, in the field of data mining and analysis, the top three venues are ACM SIGKDD, IEEE TKDE, and ACM WSDM, with \textit{h5}-indices of 69, 57, and 54, respectively. In contrast, most other venues in this field typically have \textit{h5}-indices between 10 and 20.
In light of these differences, we engage in the investigation of how different venues influence the probability that a paper contributes to the author's \hindex.
Two heuristic metrics are examined, namely (1) the average number of citations each paper in the venue collects and (2) the ratio between the number of papers in the venue with at least max-\hindex\ citations to the venue's total number of papers.
Every researcher aims to publish scientific results in well-respected journals and conferences, so our intuition is that top venues help researchers spread their scientific impact and, more specifically, to increase the citation counts of their papers, which further offers a large potential to increase their \hindices.


\subsection{Social Factors}
Previous studies show that researchers have the tendency to cite their co-authors' work \cite{Bethard:CIKM2010}.
As shown in Figure~\ref{figsub:h-numco-authors}, our preliminary results show that a researcher's \hindex\ is also positively correlated with his or her total number of collaborators/co-authors.
To explore this trend, we extract a weighted collaboration network from the dataset, where each author is denoted as a node, and each link between two nodes is connected if the researchers have collaborated with each other.
The weight of each link is defined as the frequency of collaborations.
We then extract four attributes for each node (author) from the collaboration network, including the number of co-authors (degree), the PageRank score, the average \hindex\ of co-authors, and the weighted average \hindex\ of co-authors.
For a given paper, the highest values among its authors for these four metrics are used as social factors.

\vpara{Significance.}
The factor correlations of social and venue factors are shown in Figures~\ref{figsub:cc-h-social-venue} and \ref{figsub:cc-year-social-venue}.
Both subfigures demonstrate the prominently positive correlation of venue factors.
This discovery coincides with the fact that the scientific researchers have the highly motivated tendency to publish results in top venues.
Perhaps more interestingly, the correlation between several venue factors and researchers' \hindices\ decreases as the researchers' \hindices\ increase.
This indicates that gathering into leading venues can help to initially elevate researchers' reputations, but it is the researchers' growing peer-influence itself, rather than the venue, that further increases their prominence. 
Nevertheless, compared to venue factors, social factors have only a limited correlation with whether a paper will increase its primary author's \hindex.

\subsection{Reference Factors}
The scientific impact of a scholarly work is often quantified by its respective citation count. 
In this way, the more times a publication is cited by others, the larger its assumed impact. 
Conversely, as most scientific research is undertaken by ``standing on the shoulder of giants,'' we ask whether highly cited papers actually tend to acknowledge the previous ``giants'' upon whom they stand.
Two intuitive factors are used to evaluate this question, namely (1) the ratio of a paper references that have at least max-\hindex\ citations to the paper's total number of references and (2) the average number of citations accumulated by the paper's references.

\subsection{Temporal Factors}
As fast-rising phenomena typically attract the attention of crowds more easily, a ``rising star'' in academia can attract publicity. Previous work has found that temporal information can be a powerful factor in modeling scientific impact~\cite{Bethard:CIKM2010,Yan:JCDL2012}, so it seems straightforward that the speed at which an author's \hindex\ grows should affect the rate at which her/his papers contribute to her/his \hindex.
To capture this effect, we examine the increase of authors' \hindices\ within the past three years.
Specifically, we consider four temporal factors, including the \hindex\ changes of the first author, the max-\hindex\ author, and the average change and maximum change among all authors. 
The specific definitions are shown in Table \ref{tb:factors}.

\vpara{Significance.}
From Figures~\ref{figsub:cc-h-ref-temp} and \ref{figsub:cc-year-ref-temp}, we can see that among the reference and temporal factors, \textit{R-ratio-max}, i.e., the ratio between the number of references that have accumulated at least max-\hindex\ citations and the total number of references, demonstrates the most influence on the success of \hindex\ contribution.
Other important factors are illustrated by the rising trends of both the primary author's \hindex\ and all authors' average \hindex.
However, their effects are more pronounced for the young ``rising star'' (low \hindex) than for the established, highly influential ``giants'' (high \hindex).

\subsection{Summary}
According to the correlation analysis above, we provide the following intuitions relating to academia: 
\begin{itemize}
\item A scientific researcher's authority on a topic is the most decisive factor in facilitating an increase in his or her \hindex. 
This coincides with the fact that the society fellows (e.g., ACM/IEEE fellow or membership of NAS/NAE) or lifetime honors (e.g., Turing award) are typically awarded for contributions to a topic or domain. 
\item The level of the venue in which a given paper is published is another crucial factor in determining the probability that it will contribute to its authors' \hindices.
Top venues make one outstanding and expand one's scientific impact; gradually, one's impact further helps to increase the venue's prestige.
The suggestion here lies in the every scholar's aim: \textit{Target and publish influential scientific results in top venues.}
\item People in social society often follow vogue trends.
However, publishing on an academically ``hot'' but unfamiliar topic is unlikely to further one's scientific impact, at least as measured by an increase in one's \hindex.
This reminds us that \textit{one should not turn to follow the vogue topics that are beyond his or her expertise.}
\end{itemize}  

\section{Experiments}
\label{sec:exp}


In this section, we demonstrate the predictability of whether a paper will contribute to its primary author's \hindex\ within a given timeframe.

\subsection{Experimental Setup}
The task is to predict whether the papers published in time $t$ will contribute to the \hindices\ of the authors (the max-\hindex\ author for each paper) within a given time period $\Delta t$.
For example, by setting $t=2007$, $\Delta t=5$ years, and the minimum max-\hindex\ to 10, we collect all the papers published in 2007, extract the features from the corpus before 2007, and observe whether the number of citations for each paper in 2007 is larger than or equal to the maximum \hindex\ of its authors in 2012 (the last year represented in our dataset). 

\vpara{Methods.}
As our problem is formulated as a classification task, we use a series of different classifiers, including logistic regression (LRC), random forest (RF), and bagged decision trees (BAG).
While we also generated experiments using support vector machines, na\"{\i}ve Bayes models, and neural networks, we found that their performance was very poor compared to the other three methods; thus their results have been omitted.
Due to its predictive ability, ease of implementation, and interpretability, we primarily use logistic regression to analyze factor contributions and parameter settings.

\vpara{Evaluation.}
We employ half of the instances for training and the remaining half for validation.
To quantitatively evaluate the predictability of the problem, we report the performance in terms of precision, recall, \fonescore, area under the receiver operating characteristic (AUC), and accuracy.
Furthermore, our problem can be viewed as a ranking task, namely ranking all papers of one scholar in the reverse order of probability that each paper will increase his or her \hindex.
Therefore, the precision at the top 3 results (Pre@3) and mean average precision (MAP) are also used to evaluate the performance.

\subsection{Prediction Results}

We present the predictability of whether a paper published in 2007 will contribute to the primary author's \hindex\ within five years.
The prediction is applied on the papers whose primary author's \hindex\ is at least 10 in 2007. 
The resulting set of papers contains 21,519 instances, of which 30.46\%  successfully contributed to their primary author's \hindices.
Table \ref{tb:results} lists the predictive performance of different methods.

Overall, the random guessing would achieve an \fonescore\ of 0.375 and an accuracy of 0.5.
However, by our methodology, the predictive power significantly outperforms random guessing with an \fonescore\ of over 0.776 (+107\% increase) and an accuracy of more than 0.875 (+75\% increase).
In terms of precision, recall, and AUC score, the performance is still promising.
As the three selected methods achieve similar results, logistic regression is chosen as the primary classifier to examine the remaining experiments

\begin{table}[t]
\caption{ {\bf Predictive results generated by different methods. }
$t$=2007, $\Delta t$=5 years, and \hindex\ threshold is set to 10. 
LRC --- Logistic Regression; RF --- Random Forest; BAG --- Bagged decision trees; R---Random guess with half predicted positive and half negative.
The results of support vector machine, na\"{\i}ve Bayes, and neural network models are omitted due to their poor performance on this task.
}
\label{tb:results}
\small
\centering
\renewcommand\arraystretch{1.1}
\begin{tabular}{l|l|l|l|l|l|l|l}

\hline
Method & Pre.        & Rec.     & F$_{1}$        &   AUC       &   Accu.     &    Pre@3  & MAP \\  \hline
R      & 0.305       & 0.500    & 0.375     & 0.500       &   0.500     & 0.674   & 0.522  \\
LRC    & 0.854       & 0.711    & 0.776      &   0.938    &   0.875     & 0.925         & 0.965 \\
RF     & 0.785       & 0.815    & 0.800      &   0.939    &   0.876     & 0.957         & 0.979 \\
BAG    & 0.802       & 0.821    & 0.811     &   0.951    &  0.884      & 0.950         & 0.978 \\
\hline

\end{tabular}
\vspace{-0.2cm}
\end{table}

\subsection{Factor Contribution}

\begin{figure}[t]
\centering
\includegraphics[width=3in]{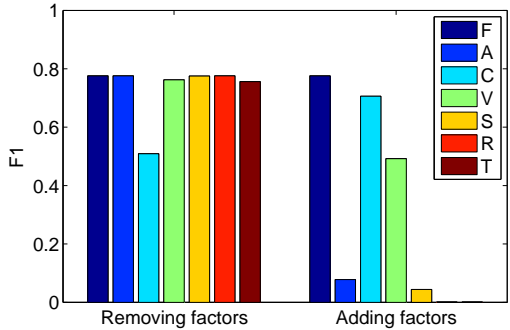}
\caption{\label{fig:factor-group-contribution}
{\bf Factor contribution analysis.} 
F denotes training a logistic regression model with a full feature set. 
A---Author factors; C---Content factors; V---Venue factors; S---Social factors; R---Reference factors; T---Temporal factors.
The left side of the figure illustrates the effects of removing each group of factors, while the right side illustrates the effects of using only each group of factors for training.
}
\vspace{-0.2cm}
\end{figure}

To predict whether a paper will increase its primary author's \hindex, we devise six groups of different factors (see \S\ref{sec:factor}) that may drive the growth of scientific impact. 
To explore the contributions of each group to the prediction task, we employ two methods: (1) we remove each group of factors and keep the remaining five to evaluate the predictive performance; and (2) we add or use only one group of factors to evaluate the predictive performance.
Figure~\ref{fig:factor-group-contribution} shows the \fonescore s for the two cases. 
The contributions of different groups of factors show a high degree of variability. 
When removing factor groups, the 30\% drop in \fonescore\ by removing the content factors indicates that they are significantly importance to this prediction task.
By contrast, the marginal decreases in performance demonstrated by all other factors imply they provide only limited contributions.
When using only one factor group, the content factors still play the most important role (0.70 \fonescore) in predicting the growth of scientific impact. 
Venue factors also achieve a reasonable performance.
The contribution analyses are consistent with the factor correlation results elaborated upon in the previous section.

We further examine the contributions of each individual factor. 
We choose the information gain ratio (IGR) to determine which of the factors are the most important. The information gain for a factor $x_k$ is the expected reduction in entropy---that is, uncertainty---achieved by learning the state of that factor.
For each factor $x_k$, then, the IGR of $x_k$, denoted $IGR(x_k)$, is simply defined as the information gain $IG(x_k)$ divided by the intrinsic value $IV(x_k)$ of $x_k$, i.e., $IGR(x_k) = IG(x_k)/IV(x_k)$. 
The higher the $IGR(x_k)$ for factor $x_k$, the greater its importance.

Table~\ref{tb:factor-igr} lists the information gain ratio and corresponding ranking of each single factor. 
Clearly, the topic authority of the primary author matters the most (\textit{C-authority-max}), followed by the authorities of the other co-authors (\textit{C-authority-ave} and \textit{C-authority-first}). 
The following positions, from the $4^{th}$ to the $7^{th}$, are held by the two venue factors and the two reference factors.
The $IGR$ of the remaining factors decrease to the next lowest order of magnitude, which indicates that they provide relatively limited contributions to the prediction task.

\begin{table}[t]
\caption{\bf{Information Gain Ratio (IGR) of each factor.}
IGR is used to decide the relevant extent of features.
Here, IGR is between 0 and 1, with higher values indicating greater relevance.
}
\label{tb:factor-igr}
\centering
\renewcommand\arraystretch{1.2}
\begin{tabular}{@{}l|l|l|r}

\hline
& Factor & IGR  & Ranking \\ \hline
\multirow{7}*{Author} & \textit{A-first-max} & 0.0028 & 21 \\
&\textit{A-num-authors}  & 0.0021 & 25 \\
& \textit{A-ave-max}     & 0.0021 & 24 \\
&\textit{A-sum-max}      & 0.0044 & 17 \\
&\textit{A-max-ratio}    & 0.0085 & 9  \\
&\textit{A-num-first}    & 0.0031 & 19 \\ 
& \textit{A-first-ratio} & 0.0049 & 16 \\

\hline

\multirow{7}*{Content} & \textit{C-popularity} & 0.0026 & 22 \\
&\textit{C-popularity-ratio} & 0.0077 & 10 \\
&\textit{C-diversity}        & 0.0013 & 26 \\
&\textit{C-novelty}          & 0.0089 & 8 \\
&\textit{C-authority-ave}    & 0.0873 & 2 \\
&\textit{C-authority-first}  & 0.0637 & 3 \\
&\textit{C-authority-max}    & 0.0963 & 1 \\

\hline

\multirow{2}*{Venue} & \textit{V-ratio-max} & 0.0473 & 4 \\ 
&\textit{V-citation} & 0.0463 & 5 \\

\hline

\multirow{4}*{Social} & \textit{S-degree} & 0.0022 & 23 \\
&\textit{S-pagerank}   & 0.0051 & 15\\
&\textit{S-h-coauthor} & 0.0073 & 11\\
&\textit{S-h-weight}   & 0.0052 & 14\\

\hline

\multirow{2}*{Reference} & \textit{R-ratio-max} & 0.0127 & 6 \\
&\textit{R-citation} & 0.0105 & 7 \\

\hline

\multirow{4}*{Temporal} & \textit{T-ave-h} & 0.0039 & 18 \\ 
&\textit{T-max-h}   & 0.0056 & 13 \\
&\textit{T-h-first} & 0.0028 & 20 \\ 
&\textit{T-h-max}   & 0.0056 & 12 \\

\hline

\end{tabular}
\vspace{-0.2cm}
\end{table}

\subsection{Predictability of Papers with Different Constraints}

Our experimental results provide evidence of the predictability of whether one paper will contribute to the \hindex\ of it primary author within five years.
Two intuitive questions naturally arise:
First, is a primary author with a high or a low \hindex\ more predictable?
Second, is a paper more predictable given a long or short timeframe?
To answer these questions, we investigate the predictability of papers conditioned on the primary author's \hindex\ and the length of the given timeframe $\Delta t$.

\begin{figure}[t]
\centering
\includegraphics[width=3in]{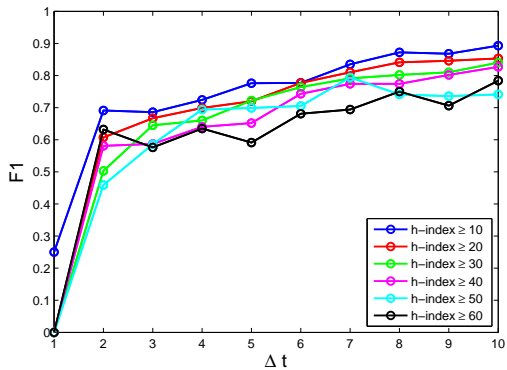}
\caption{\label{fig:varying-para}
{\bf Predictive performance by varying the constraints.} 
}
\vspace{-0.2cm}
\end{figure}

Figure~\ref{fig:varying-para} shows the predictive performance with different constraints. 
On one hand, it suggests that our prediction task is more difficult for papers with a high \hindex\ primary author.
Recall that the higher the primary author's \hindex, the higher the local threshold for this paper during classification.
Intuitively, then, as we consider authors with higher \hindices, it becomes increasingly difficult for any particular paper to reach the defined local threshold, which makes the corresponding prediction task increasingly non-trivial.
On the other hand, the rising lines as the increase of $\Delta t$ imply that our prediction task is easier when given a longer timeframe.
Intuitively, papers can accrue more citations as time goes on, during which time the authors' influence may increase, which may further compound the rate at which citations accrue. 
In this way, it becomes quite likely that the papers will contribute to their respective authors' \hindices\ if simply given a long enough timeframe.

Overall, the task of determining which papers will increase one' \hindex\ is more predictable when conducted over a relatively long timeframe for those who have relatively low \hindices.

\subsection{Case Studies}

We illustrate several case studies to demonstrate the effectiveness of our methodology for quantifying scientific impact. Our case studies include two researchers and two data mining conferences, and conclude with the self-prediction of \textit{this} paper.

\begin{figure}[t]
\centering
\subfigure[\scriptsize $\#$citations of two authors' papers.]{
\hspace{-0.2in}
\label{figsub:case-hanpei}
\includegraphics[width=1.75in]{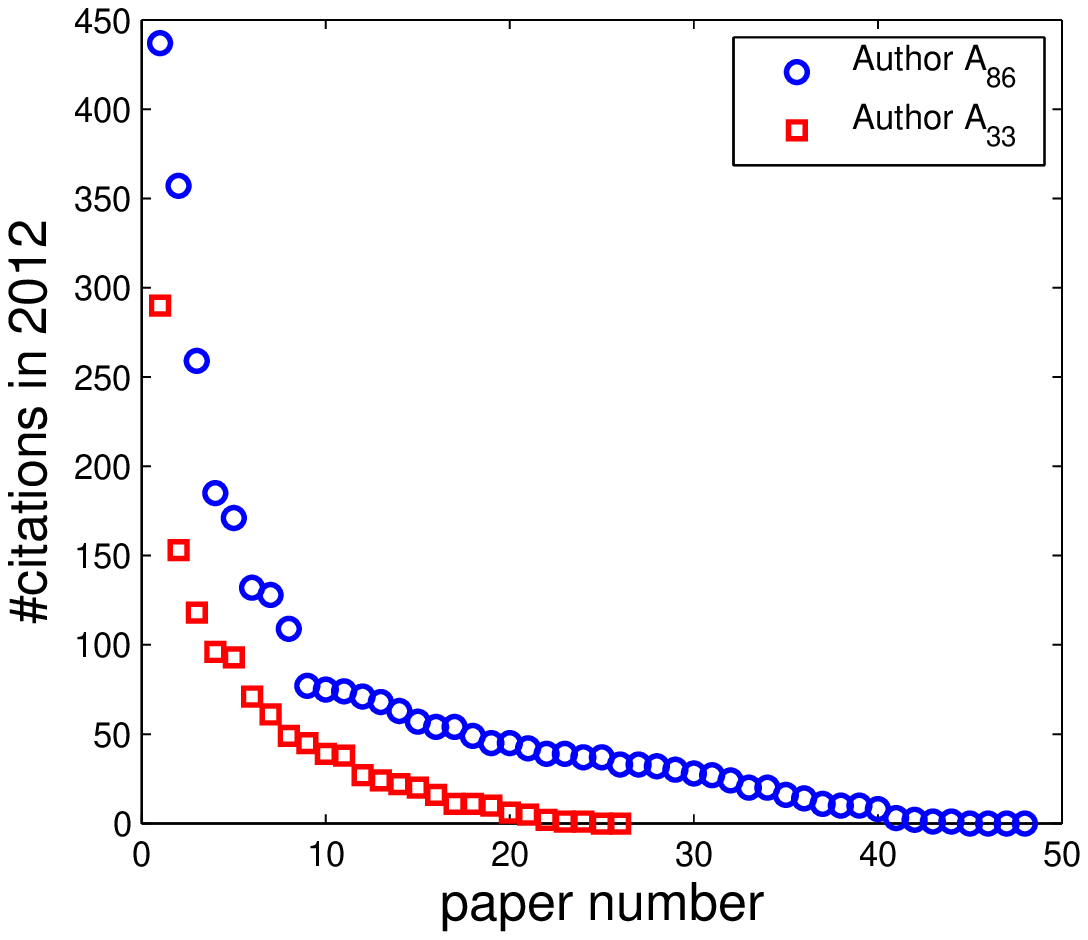}
}
\hspace{-0.1in}
\subfigure[\scriptsize $\#$citations of two venues' papers.]{
\label{figsub:case-kddicdm}
\includegraphics[width=1.75in]{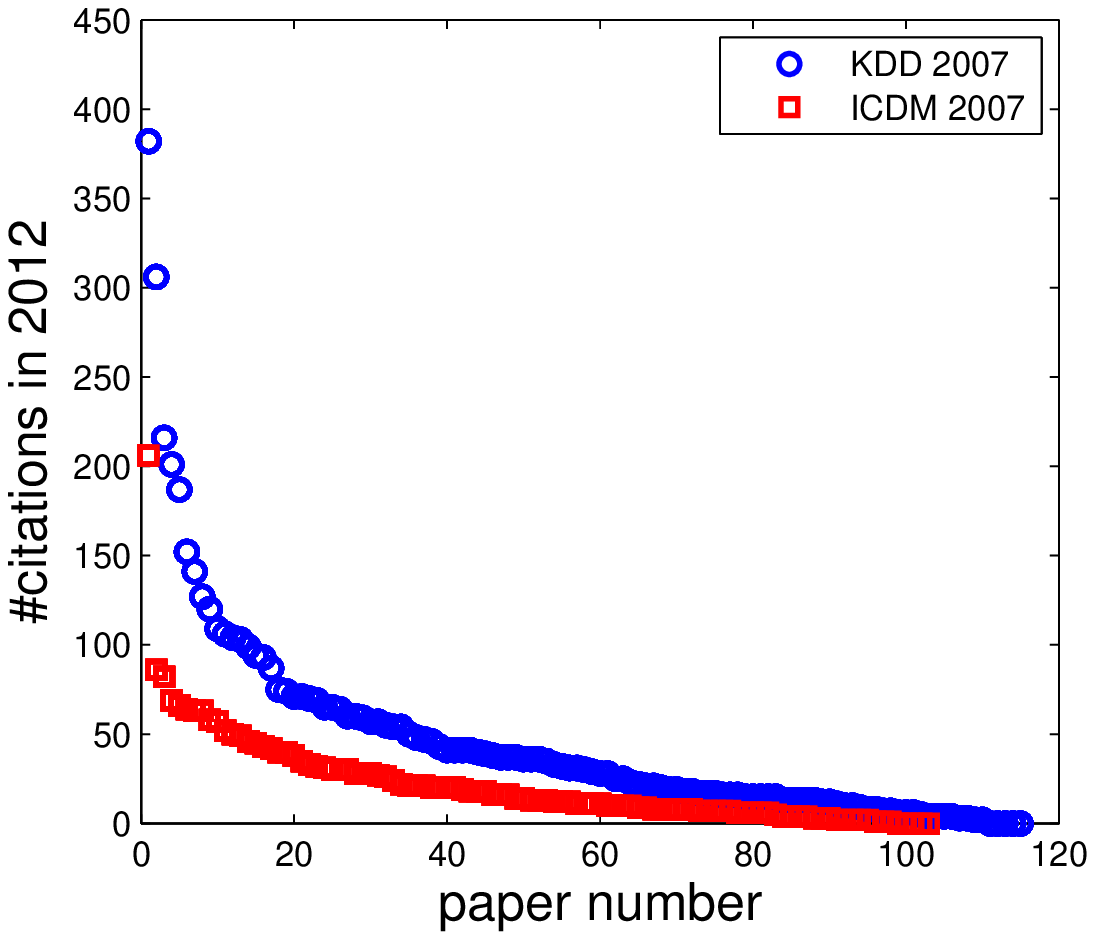}
\hspace{-0.2in}
}
\caption{\label{fig:case-hanpei-kddicdm}
{\bf Case studies of citation counts in 2012 for papers published in 2007.} 
In 2012, 8 of $A_{86}$'s 48 papers and 9 of $A_{33}$'s 26 papers published in 2007 obtained $\#$citations over 86 and 33 ($A_{86}$'s and $A_{33}$'s \hindices\ in 2007), respectively. 
}
\vspace{-0.2cm}
\end{figure}

\vpara{Two anonymous scholars.}
We choose two anonymous scholars, $A_{86}$ and $A_{33}$, who in 2007 had substantially different \hindices\ (86 and 33, respectively).
We examine the difference in the predictability of whether the papers published by each author in 2007 would contribute to his or her respective \hindex\ by 2012. 
From Figure~\ref{figsub:case-hanpei}, we can observe the skewed heavy-tailed distributions for the number of citations their papers published in 2007 have obtained by 2012. 
Back in 2007, $A_{86}$'s and $A_{33}$'s \hindices\ were 86 and 33, respectively.
In 2012, 8 of $A_{86}$'s 48 publications published in 2007 have obtained at least 86 citations and 9 of $A_{33}$'s 26 publications have obtained at least 33. 
In other words, while the ratio of the papers that have contributed to $A_{86}$'s \hindex\ is relatively low (8/48), this is actually a result of the researcher's high \hindex, which corresponds to a large scientific impact.
In contrast, $A_{33}$'s relatively low \hindex\ compared to $A_{86}$ makes the ratio of papers that contribute to $A_{33}$'s \hindex\ higher than those that contribute to $A_{86}$'s. 
Table~\ref{tb:hanpei} shows the predictive results of whether their papers in 2007 would contribute to their \hindices\ in 2012, from which we make two observations. 
First, the performance of the model for both researchers is much higher than random guessing with respect to \fonescore, AUC, and accuracy. 
Second, the overall performance for $A_{33}$ is significantly higher than that for $A_{86}$, which suggests a greater predictability for researchers with relatively lower \hindices.

\vpara{KDD and ICDM.}
Using ACM SIGKDD and IEEE ICDM as examples, we present the effects that different conferences have on spreading scientific impact and contributing to authors' \hindices. 
Currently, KDD and ICDM are considered two of the world's premier research conferences in data mining. 
Back in 2007, ICDM was in its seventh year, while KDD was in its thirteenth year. 
Figure~\ref{figsub:case-kddicdm} plots the number of citations in 2012 for all papers published in these two venues in 2007. 
In our dataset, we observe that KDD'07 and ICDM'07 papers contribute to their primary authors' \hindices\ in different proportions. 
We further examine which conference has papers that are, in terms of their contribution to the \hindices\ of their respective authors, more predictable. 
This comparison is illustrated by Table~\ref{tb:kddicdm}. 
The prediction results for KDD'07 papers are generally better than ICDM'07 in terms of recall and and \fonescore, and are competitive with ICDM'07 in terms of precision, AUC, and accuracy. 
Overall, both venues demonstrate promising predictability in the spread of the scientific impact of their respective publications.


\vpara{Self-prediction for \textit{this} paper.}
Finally, we perform a self-prediction on whether \textit{this} paper will contribute to its primary author's \hindex\ by 2019. 
Keeping in mind that using the methodology employed in this paper would technically require unavailable training data (i.e., data composed of publications from 2014 and their citation counts in 2019), we instead revert to using the logistic regression model trained on the data from 2007 to generate this prediction.
To generate features for this paper, we insert this paper into the 2012 data and make the assumption that it was accepted by the venue to which we have submitted at that time. 
The author factors, for example, are extracted from their respective author profiles circa 2012. 
With the approximate model and features, we estimate that the probability that this paper will contribute to the \hindex\ of the author with highest \hindex\ within five years is around 76\%. With a pinch of optimism, we leave the evaluation of the fidelity of this prediction as an exercise for future readers.

\begin{table}[t]
\caption{ {\bf Predictive results for two anonymous authors. }
$t$=2007 and $\Delta t$=5 years. Pre@k: $k$=8 ($A_{86}$); $k$=9 ($A_{33}$).
}
\label{tb:hanpei}
\small
\centering
\renewcommand\arraystretch{1.1}
\begin{tabular}{l|l|l|l|l|l|l|l}

\hline
Authors     & Pre.        & Rec.     & F$_{1}$        &   AUC       &   Accu.     &    Pre@k  & MAP \\  \hline
$A_{86}$    & 0.500       & 0.375    & 0.429      &   0.584    &   0.833     & 0.375    & 0.346 \\
$A_{33}$    & 1.000       & 0.667    & 0.800     &   0.856    &  0.885      & 0.667      & 0.849 \\
\hline

\end{tabular}
\vspace{-0.2cm}
\end{table}

\begin{table}[t]
\caption{ {\bf Predictive results for KDD'07 and ICDM'07 papers. }
$t$=2007, $\Delta t$=5 years and logistic regression is used. In our dataset, KDD'07 and ICDM'07 papers contribute to their primary authors' \hindices\ in different proportions. 
}
\label{tb:kddicdm}
\small
\centering
\renewcommand\arraystretch{1.1}
\begin{tabular}{l|l|l|l|l|l}

\hline
Venues         & Pre.        & Rec.     & F$_{1}$        &   AUC       &   Accu.      \\  \hline
{KDD'07}    & 0.800       & 0.889    & 0.842      &   0.884    &   0.818     \\
{ICDM'07}    & 0.842       & 0.593    & 0.696     &   0.886    &  0.825     \\
\hline

\end{tabular}
\vspace{-0.2cm}
\end{table} 

\section{Related work}
\label{sec:related}

Scientific impact modeling is being extensively explored and has become an important and popular research topic, particularly since the rise of two recent, successive \textit{Science} papers in 2013~\cite{Uzzi:Science13,Wang:Science13}.
The study of scientific impact by scholarly researchers offers the potential to help scholars more effectively disseminate their work and expand their scientific influence.

Traditionally, the number of citations has been widely used as a measurement of scientific impact for both individual papers and solitary scientific researchers.
Several practical metrics are designed based on the number of citations to reflect scientific impacts.
Garfield proposed the idea of impact factor for indexing and evaluating the quality of journals~\cite{Garfield:Science1955}.
Recently, Hirsch proposed the \hindex, which is able to measure both the productivity and popularity of researchers~\cite{Hirsch:05}. 
In this setting, on the one hand scholars aim to publish results in high-impact journals to collect more citations and increase their \hindices, while on the other hand the journals attempt to accept solid and potentially influential work to improve their impact factors.
Accordingly, both impact factor and \hindex\ successfully characterize the motivation and behaviors of the scientific community.

Besides measuring scientific impact, a large body of work has been focused on the prediction of scientific impact. 
The 2003 ACM SIGKDD Cup introduced a competition focused around citation count prediction~\cite{KDDCUP:03}, with the task of estimating the number of citations of one paper given its previous number of citations.
Following this, many efforts have been made to predict the number of future citations for scholarly work.
Castillo et al. studied the correlation between author reputation and citations~\cite{Castillo:2007}.
Yan et al. examined a series of features important to future citations~\cite{Yan:JCDL2012,Yan:CIKM2011}.
Wang et al. uncovered basic mechanisms that govern scientific impact, which has the power to quantify and predict citation counts~\cite{Wang:Science13,Shen:AAAI14}.
However, the effectiveness of such predictions is fundamentally limited by the power law distribution of citations.
Herein we (re)define the impact prediction problem by addressing a question of interest to many academic researchers, namely: ``will this paper increase my \hindex?''
The major difference between our work and previous work lies in that, rather than solving a regression task in a highly skewed environment, we address the problem by generating a local threshold (the author's \hindex) for each paper's future citation count.

Our work is also related to other mining tasks in academic data such as citation recommendation~\cite{Ren:KDD14,Tang:2009citation,XiaoYu:SDM12}, topic influence~\cite{Liu:CIKM2010,Tang:09KDD}, collaboration prediction~\cite{Sun:WSDM2012,Wang:10KDD}, and analysis of citation networks~\cite{Smyth:ICML11} and academic social networks~\cite{Tang:08KDD}. 
Furthermore, as the formalization of our predictive task is partly inspired by the cascade growth prediction problem~\cite{Cheng:WWW14}, the prediction of scientific impact has relations with predicting the popularity~\cite{Hong:WSDM13,Pinto:WSDM13,Ahmed:WSDM13} of online ``paper'' (e.g., tweet, video, photo) in social media. 

\section{Conclusion}
\label{sec:conclusion}

In this paper, we examine scientific impact by formalizing a novel problem that can be reduced to the following question: will a paper contribute to its primary author's \hindex\ within a given timeframe? 
Previous, rich work on predicting the number of citations of individual papers is fundamentally limited by the heavy-tailed distribution of citation counts.
Our problem definition differs by offering a great potential to quantify scientific impact on the interplay between scientific researchers and publications.
We examine six groups of factors from different perspectives that may drive a paper to obtain enough citations to contribute to its primary author's \hindex. 

By experimenting on a large-scale public academic dataset, we find that two factors---(1) the authors' authority on the publication topic and (2) the publication venue---play the most decisive roles in determining whether a paper will contribute to its primary author's \hindex.
Surprisingly, we notice that the popularity of the publication topic and the co-authors' influence are not correlated to the prediction target. 
Our study also demonstrates a greater than 87.5\% potential predictability for whether a paper will contribute to its primary author's \hindex\ within five years.
Overall, our findings unveil mechanisms for quantifying scientific impact and provide concrete suggestions to researchers for better expanding their scientific influence and, ultimately, for more effectively ``standing on the shoulders of giants.''

Notwithstanding the extensive and promising results of the present work, there is still much room left for future work.
First, while this work is conducted only on literature from computer science, it is necessary to examine the observed patterns in other scientific disciplines, such as physics, mathematics, biology, and so on.
Second, since authors' \hindices\ evolve within the prediction timeframe, it would be natural to design methodologies that could capture the co-evolution of authors' \hindices\ and citation counts.

\vpara{Acknowledgments.}\small
This work is supported by the Army Research Laboratory under Cooperative Agreement Number W911NF-09-2-0053, the U.S. Air Force Office of Scientific Research (AFOSR) and the Defense Advanced Research Projects Agency (DARPA) grant $\#$FA9550-12-1-0405, and the National Science Foundation (NSF) Grant OCI-1029584.

\normalsize

\balance
\small
\bibliographystyle{abbrv}
\bibliography{references-full}  

\begin{thebibliography}{10}

\bibitem{Ahmed:WSDM13}
M.~Ahmed, S.~Spagna, F.~Huici, and S.~Niccolini.
\newblock A peek into the future: Predicting the evolution of popularity in
  user generated content.
\newblock In {\em WSDM '13}, pages 607--616. ACM, 2013.

\bibitem{Bethard:CIKM2010}
S.~Bethard and D.~Jurafsky.
\newblock Who should {I} cite: Learning literature search models from citation
  behavior.
\newblock In {\em CIKM '10}, pages 609--618. ACM, 2010.

\bibitem{Blei:03}
D.~M. Blei, A.~Y. Ng, and M.~I. Jordan.
\newblock Latent {D}irichlet allocation.
\newblock {\em JMLR}, 3:993--1022, 2003.

\bibitem{Castillo:2007}
C.~Castillo, D.~Donato, and A.~Gionis.
\newblock Estimating the number of citations using author reputation.
\newblock In {\em SPIRE '07}, pages 107--117. Springer, 2007.

\bibitem{Cheng:WWW14}
J.~Cheng, L.~Adamic, P.~A. Dow, J.~M. Kleinberg, and J.~Leskovec.
\newblock Can cascades be predicted?
\newblock In {\em WWW '14}, pages 925--936, 2014.

\bibitem{Garfield:Science1955}
E.~Garfield.
\newblock Citation indexes for science: A new dimension in documentation
  through association of ideas.
\newblock {\em Science}, 122(3159):108--111, 1955.

\bibitem{KDDCUP:03}
J.~Gehrke, P.~Ginsparg, and J.~M. Kleinberg.
\newblock Overview of the 2003 kdd cup.
\newblock {\em SIGKDD Explorations}, 5(2):149--151, 2003.

\bibitem{Hirsch:05}
J.~E. Hirsch.
\newblock An index to quantify an individual's scientific research output.
\newblock {\em Proceedings of the National Academy of Sciences},
  102(46):16569--16572, 2005.

\bibitem{Hong:WSDM13}
L.~Hong, A.~S. Doumith, and B.~D. Davison.
\newblock Co-factorization machines: Modeling user interests and predicting
  individual decisions in {T}witter.
\newblock In {\em WSDM '13}, pages 557--566. ACM, 2013.

\bibitem{Kleinberg:99}
J.~M. Kleinberg.
\newblock Authoritative sources in a hyperlinked environment.
\newblock {\em Journal of the ACM}, 46(5):604--632, 1999.

\bibitem{Kleinberg:STOC11}
J.~M. Kleinberg and S.~Oren.
\newblock Mechanisms for (mis)allocating scientific credit.
\newblock In {\em STOC '11}, pages 529--538. ACM, 2011.

\bibitem{Kullback:51}
S.~Kullback and R.~A. Leibler.
\newblock On information and sufficiency.
\newblock {\em Annals of Machematical Statistics}, 22(1):79--86, 1951.

\bibitem{Liu:CIKM2010}
L.~Liu, J.~Tang, J.~Han, M.~Jiang, and S.~Yang.
\newblock Mining topic-level influence in heterogeneous networks.
\newblock In {\em CIKM '10}, pages 199--208. ACM, 2010.

\bibitem{Pinto:WSDM13}
H.~Pinto, J.~M. Almeida, and M.~A. Gon\c{c}alves.
\newblock Using early view patterns to predict the popularity of youtube
  videos.
\newblock In {\em WSDM '13}, pages 365--374. ACM, 2013.

\bibitem{Radicchi:PNAS08}
F.~Radicchi, S.~Fortunato, and C.~Castellano.
\newblock Universality of citation distributions: Toward an objective measure
  of scientific impact.
\newblock {\em PNAS}, 2008.

\bibitem{Ren:KDD14}
X.~Ren, J.~Liu, X.~Yu, U.~Khandelwal, Q.~Gu, L.~Wang, and J.~Han.
\newblock Clus{C}ite: Effective citation recommendation by information
  network-based clustering.
\newblock In {\em KDD '14}, 2014.

\bibitem{Shen:PNAS14}
H.-W. Shen and A.-L. Barab{\'a}si.
\newblock Collective credit allocation in science.
\newblock {\em PNAS}, 2014.

\bibitem{Shen:AAAI14}
H.-W. Shen, D.~Wang, C.~Song, and A.-L. Barab{\'a}si.
\newblock Modeling and predicting popularity dynamics via reinforced poisson
  processes.
\newblock In {\em AAAI '14}, 2014.

\bibitem{strathern1997improving}
M.~Strathern.
\newblock Improving ratings: audit in the {B}ritish university system.
\newblock {\em European Review}, 5(03):305--321, 1997.

\bibitem{Sun:WSDM2012}
Y.~Sun, J.~Han, C.~C. Aggarwal, and N.~V. Chawla.
\newblock When will it happen?: Relationship prediction in heterogeneous
  information networks.
\newblock In {\em WSDM '12}, pages 663--672. ACM, 2012.

\bibitem{Tang:09KDD}
J.~Tang, J.~Sun, C.~Wang, and Z.~Yang.
\newblock Social influence analysis in large-scale networks.
\newblock In {\em KDD '09}, pages 807--816, 2009.

\bibitem{Tang:2009citation}
J.~Tang and J.~Zhang.
\newblock A discriminative approach to topic-based citation recommendation.
\newblock {\em Advances in Knowledge Discovery and Data Mining}, pages
  572--579, 2009.

\bibitem{Tang:08KDD}
J.~Tang, J.~Zhang, L.~Yao, J.~Li, L.~Zhang, and Z.~Su.
\newblock Arnetminer: Extraction and mining of academic social networks.
\newblock In {\em KDD '08}, pages 990--998, 2008.

\bibitem{Uzzi:Science13}
B.~Uzzi, S.~Mukherjee, M.~Stringer, and B.~Jones.
\newblock Atypical combinations and scientific impact.
\newblock {\em Science}, 342(6157):468--472, 2013.

\bibitem{Smyth:ICML11}
D.~Vu, A.~Asuncion, D.~Hunter, and P.~Smyth.
\newblock Dynamic egocentric models for citation networks.
\newblock In {\em ICML '11}, pages 857--864, 2011.

\bibitem{Wang:10KDD}
C.~Wang, J.~Han, Y.~Jia, J.~Tang, D.~Zhang, Y.~Yu, and J.~Guo.
\newblock Mining advisor-advisee relationships from research publication
  networks.
\newblock In {\em KDD '10}, pages 203--212, 2010.

\bibitem{Wang:Science13}
D.~Wang, C.~Song, and A.-L. Barab{\'a}si.
\newblock Quantifying long-term scientific impact.
\newblock {\em Science}, 342(6154):127--132, 2013.

\bibitem{Yan:JCDL2012}
R.~Yan, C.~Huang, J.~Tang, Y.~Zhang, and X.~Li.
\newblock To better stand on the shoulder of giants.
\newblock In {\em JCDL '12}, pages 51--60. ACM, 2012.

\bibitem{Yan:CIKM2011}
R.~Yan, J.~Tang, X.~Liu, D.~Shan, and X.~Li.
\newblock Citation count prediction: Learning to estimate future citations for
  literature.
\newblock In {\em CIKM '11}, pages 1247--1252. ACM, 2011.

\bibitem{XiaoYu:SDM12}
X.~Yu, Q.~Gu, M.~Zhou, and J.~Han.
\newblock Citation prediction in heterogeneous bibliographic networks.
\newblock In {\em SDM '12}, pages 1119--1130, 2012.

\bibitem{Zhang:07DASFAA}
J.~Zhang, J.~Tang, and J.~Li.
\newblock Expert finding in a social network.
\newblock In {\em DASFAA '07}, pages 1066--1069, 2007.

\end{thebibliography}

\end{document}